\documentclass[12pt,a4paper]{article}
\usepackage{latexsym}
\usepackage{epsf}
\font\mybb=msbm10 at 12pt
\def\bb#1{\hbox{\mybb#1}}
\def\Z {\bb{Z}}
\def\R {\bb{R}}

\font\mybb=msbm10 at 10pt
\def\bb#1{\hbox{\mybb#1}}
\def\Z {\bb{Z}}
\def\R {\bb{R}}

\makeatletter
\@addtoreset{equation}{section}
\makeatother

\def\unit{\hbox to 3.3pt{\hskip1.3pt \vrule height 7pt width .4pt \hskip.7pt
\vrule height 7.85pt width .4pt \kern-2.4pt
\hrulefill \kern-3pt
\raise 4pt\hbox{\char'40}}}

\def\be{\begin{equation}}
\def\ee{\end{equation}}
\def\bea{\begin{eqnarray}}
\def\eea{\end{eqnarray}} 

\def\ba{\begin{array}}
\def\ea{\end{array}}

\def\part{\partial}

\def \alfa {2 \pi \alpha'}
\def \ikC {({i}_{\hat k} {\hat C})}

\begin{document}
\begin{flushright}
\footnotesize
\footnotesize
UG-15/99\\
SPIN-99/18 \\
August, $1999$
\normalsize
\end{flushright}

\begin{center}

\vspace{.6cm}
{\LARGE {\bf Exotic Branes and Nonperturbative Seven Branes}}

\vspace{.9cm}


{\bf Eduardo Eyras}

\vspace{.1cm}

{
{\it Institute for Theoretical Physics,
University of Groningen \\
Nijenborgh 4, 9747 AG Groningen, The Netherlands}\\
{\tt E.A.Eyras@phys.rug.nl}
}

\vspace{.3cm}

{and}

\vspace{.3cm}

{\bf Yolanda Lozano}

\vspace{.1cm}

{
{\it Spinoza Institute,
University of Utrecht\\
Leuvenlaan 4, 3508 TD Utrecht, The Netherlands}\\
{\tt Y.Lozano@phys.uu.nl}
}

\vspace{.2cm}


\vspace{.2cm}

\vspace{.8cm}


{\bf Abstract}

\end{center}
\begin{quotation}

\small

We construct the effective action of certain exotic branes in the
Type II theories which are not predicted by their spacetime supersymmetry
algebras. We analyze in detail the case of the NS-7B brane,
S-dual to the D7-brane, and connected by T-duality to other exotic
branes in Type IIA: the KK-6A brane and the KK-8A brane (obtained by
reduction of the M-theory Kaluza-Klein monopole and M9-brane,
respectively).
The NS-7B brane carries charge with respect to the
S-dual of the RR 8-form, which we identify as a non-local combination of the 
electric-magnetic duals of the axion and the dilaton. The study
of its effective action agrees with previous results in the literature
showing that it transforms as an $SL(2,\Z)$ triplet together 
with the D7-brane. We discuss
why this brane is not predicted by the Type IIB spacetime
supersymmetry algebra. In particular we show that the modular transformation
relating the D7 and NS-7B brane solutions can be undone by a simple
coordinate transformation in the two dimensional transverse space,
equivalent to choosing a different region to parametrize the
$SL(2,\Z)$ moduli space.
We discuss a similar relation between the D6 and KK-6A branes and the
D8 and KK-8A branes.

\end{quotation}

\vspace{1cm}

\newpage

\pagestyle{plain}


\newpage
\section{Introduction}


It is by now well-known that the BPS spectra of the Type II
and M- theories can be obtained by analyzing the central charges that
occur in the corresponding supersymmetry algebra \cite{H4,T}.
This analysis predicts a 0-brane (the eleven dimensional pp-wave), a 2-brane, 
a 5-brane, a 6-brane (the Kaluza-Klein monopole) and a 9-brane in
M-theory. These branes give rise to the brane-scan of the
Type IIA theory, namely the 0-brane, the fundamental string, 
the NS-5A brane, D$p$-branes
with $p$ even, $p=0,\dots ,8$, the Kaluza-Klein monopole 
(or KK-5A brane) and the NS-9A brane.
All these branes are predicted as well by the Type IIA supersymmetry
algebra and other than the NS-9A brane have been extensively 
studied in the literature.
The NS-9A brane is a spacetime-filling brane which may play a
role in the construction of the Heterotic string as a
non-perturbative orientifold of
the Type IIA theory (see \cite{BEHHLvdS} for some results in this direction). 

{}From the Type IIA brane-scan, via T-duality, one obtains the 
brane-scan of the Type IIB 
theory, also predicted by its spacetime supersymmetry algebra. It
consists on the 0-brane, the fundamental string, 
the NS-5B brane, D$p$-branes with
$p$ odd, $p=-1,1,\dots ,9$, the Kaluza-Klein monopole 
(or KK-5B brane) and the NS-9B brane. These branes occur as multiplets
of the SL(2,\Z) symmetry of the theory. 
Namely the (F1,D1), (NS5B,D5)
and (NS9B,D9) branes occur as doublets and the D3-brane, D7-brane and
KK-monopole as singlets.
The D9-brane and NS-9B brane are spacetime-filling branes which 
are charged with respect
to the RR 10-form and its S-dual, a NS-NS 10-form \cite{BRJO}, 
respectively.
It is well-known
that D9-branes play a role in the construction of the Type I theory 
as an orientifold of Type IIB. Similarly, it was argued in 
\cite{H3,BEHHLvdS} that NS-9B branes play a role in the construction
of the Heterotic SO(32) theory as a non-perturbative orientifold 
of Type IIB. In this
case 32 NS-9B branes cancel the charge of the orientifold fixed plane
associated to the S-dual operation of worldsheet parity reversal. 

It has been pointed out however that there exist some brane 
solutions in the
Type IIA and Type IIB theories which are not predicted by their
spacetime supersymmetry algebras. These branes have been encountered in
different contexts. On the one hand they are required in order to 
fill up multiplets of BPS states in representations of the
U-duality group of M-theory on a d-torus \cite{EGKR,BOL,H1,OP}. 
On the other hand they are
predicted by duality from branes that do occur as central
charges in the spacetime supersymmetry algebras. 

In particular it is well-known that the
Kaluza-Klein monopole of M-theory gives rise to the D6-brane after
dimensional reduction along its Taub-NUT direction and to the Type IIA
KK-monopole after double dimensional reduction. The third possibility,
reducing along a transverse direction different from the Taub-NUT
direction, gives rise to a 6-brane in the Type IIA theory which
contains as well a Taub-NUT fiber direction 
\cite{EGKR,BOL} (we will denote it as 
a KK-6A brane) but which is not predicted by the Type IIA
spacetime supersymmetry algebra.
The corresponding supergravity solution is described by a harmonic
function on ${\R}^2\times S^1$ and therefore is
logarithmically divergent \cite{BOL}.
This fact was indicated as an explanation for its non-occurrence in the
Type IIA spacetime supersymmetry algebra \cite{H1}, given that
these algebras
are realized as asymptotic symmetries, with flat asymptotic metric. 
However it is still striking that
the D7-brane, being as well logarithmically divergent, 
or the D8-brane, linearly divergent, 
have associated
central charges in the Type IIB and Type IIA 
spacetime supersymmetry algebras.

A second brane in the Type IIA theory not predicted either
by its spacetime supersymmetry algebra occurs when double dimensionally 
reducing the M9-brane. A single M9-brane has associated a Killing direction 
which is interpreted as a worldvolume direction, and which appears
gauged in the effective action \cite{H1,BvdS,EL1}.
Reduction along this direction gives the D8-brane. 
On the other hand, reducing along the transverse direction we obtain the
NS-9A brane mentioned previously, which contains a gauged direction
also interpreted as a worldvolume direction.
The third possibility consists in reducing along a 
worldvolume direction, different from the Killing direction. This predicts
a new 8-brane in the Type IIA theory with a gauged direction
in its worldvolume, which we denote as a KK-8A brane.

\begin{figure}[!ht]
\begin{center}
\leavevmode
\epsfxsize= 13cm
\epsfysize= 6cm
\epsffile{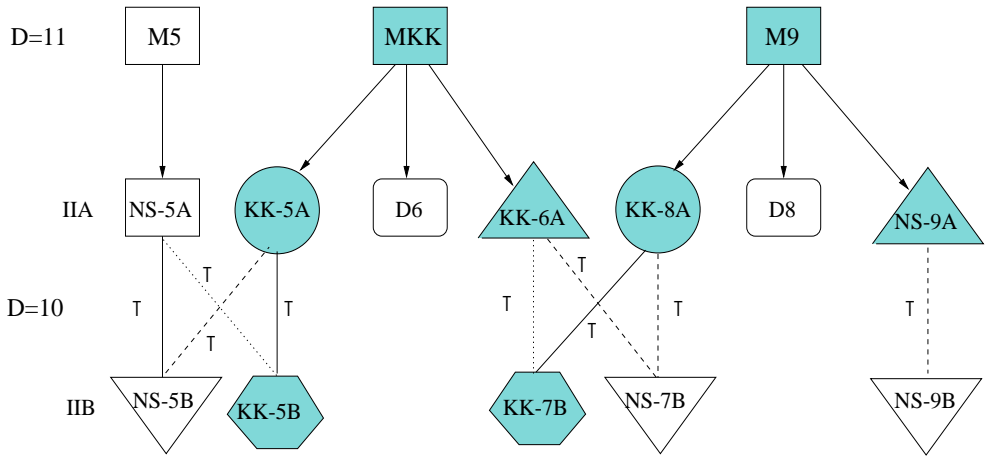}
\caption{\small {\bf Exotic branes and T-duality.}
In this figure we show the relation under T-duality of the different branes
considered in this article . We also include the 5-branes and
Kaluza-Klein monopoles for comparison of these relations,
as well as the M-theory origin of the type IIA branes.
With respect to the type IIA branes,
we indicate a T-duality along a worldvolume direction by a solid line, 
a T-duality along a Killing direction with a dashed line, 
and a T-duality along a transverse direction with
a dotted line.}
\label{grafico1}
\end{center}
\end{figure}

The KK-6A and KK-8A branes are related by T-duality to a new 7-brane
in Type IIB (see figure \ref{grafico1}) which is also predicted by the
S-duality symmetry of the theory \cite{PT,EL1}. 
Performing an S-duality transformation in the worldvolume
effective action of a D$p$-brane in Type IIB:

\begin{equation}
S=\int d^{p+1}\xi e^{-\phi} \sqrt{|{\rm det}(g+(2\pi\alpha^\prime)
{\cal F})|}\, ,
\end{equation}

\noindent with ${\cal F}=2\partial b+\frac{1}{2\pi\alpha^\prime}
B^{(2)}$ and $B^{(2)}$ the (pullback of the) NS-NS 2-form, 
we obtain a $p$-brane with effective action

\begin{equation}
\label{Sduals}
S=\int d^{p+1}\xi e^{-(\frac{p-1}{2})\phi}
(1+e^{2\phi}(C^{(0)})^2)^{\frac{p-3}{4}}\sqrt{|{\rm det}
(g+\frac{(2\pi\alpha^\prime)e^{\phi}}{\sqrt{1+e^{2\phi}(C^{(0)})^2}}
{\tilde {\cal F}})|}\, ,
\end{equation}

\noindent where ${\tilde {\cal F}}=2\partial c^{(1)}+
\frac{1}{2\pi\alpha^\prime}C^{(2)}$, $C^{(2)}$ is the (pullback
of the) RR 2-form and $b\rightarrow -c^{(1)}$ under S-duality.
For $p=1$ (\ref{Sduals})
reduces to the Nambu-Goto action of the
F-string after dualizing the 1-form $c^{(1)}$ into
a constant scalar\footnote{The value of this scalar fixes the number
of F-strings after quantization.}. For $p=3$ it does not describe
a new 3-brane since worldvolume duality of $c^{(1)}$ yields the
effective action of the D3-brane, showing that this brane is invariant
under S-duality \cite{GGTL}. 
For $p=5,9$ we obtain the kinetic terms of the 
NS-5B and NS-9B brane effective actions\footnote{See \cite{EJL} and
\cite{BEHHLvdS}, respectively.},
which form doublets with the D5 and
D9 branes respectively.
However, for $p=7$ it describes
a 7-brane which is not predicted by the Type IIB spacetime supersymmetry
algebra and which we will denote as a NS-7B brane.
This brane cannot be related to the D7-brane by worldvolume duality.
The corresponding supergravity solution, 
as well as its connections via T-duality with
the KK-6A and KK-8A branes in Type IIA, has been studied in \cite{PT}.
In this reference it is also shown that the NS-7B
brane can be cast into a triplet of $SL(2,\Z)$ 
together with the D7-brane. The third charge is the magnetic dilaton
charge carried by the 7-brane, and is related to the other two
charges upon quantization. Therefore a generic 7-brane can
be consistently denoted by two integers. The values of these integers
are further restricted by the characterization of the 7-branes by their
monodromy matrices or, equivalently, by the type of strings that
can end on them \cite{DL}.

It is the purpose of this paper to study in some more detail
these ``exotic'' branes in the
Type IIA and Type IIB theories which are not predicted by their spacetime
supersymmetry algebras, but are however needed
in order to fill up the multiplets of BPS states predicted by
the $E_7(\Z)$, $E_8(\Z)$ U-duality groups of M-theory on $T^7$, $T^8$
\cite{EGKR,BOL,H1,OP}.
All the exotic branes that we analyze share
the property that the tension scales with $1/g_s^3$, with $g_s$ the
string coupling constant, so that they are non-perturbative, i.e.
they do not have an obvious interpretation in weakly coupled string
theory. 

We construct the worldvolume effective actions by means of duality
transformations on ordinary branes. 
We mainly focus on the NS-7B brane, and we see
from its worldvolume effective action that 
it does not form a doublet with the D7-brane. We identify the 8-form
to which this brane is minimally coupled as a non-local combination
of the electric-magnetic duals of the axion and the dilaton. 
This potential needs to be introduced in order to study the
behavior of the D7-brane under S-duality.
Its field strength however is a local combination of 
the 9-form field strengths associated to the duals of the axion and
dilaton, so that a dual Type IIB supergravity, written as a function
of the dual 9-form field strengths, will not depend on it
\cite{D-Lechner-Tonin, CGNSW}. 
At the level of the solutions we see that the D7 and
NS-7B branes are related by a modular transformation, but that this
transformation
can be undone by a change of coordinates in the transverse
space. This means that both branes represent the same physical object
in different coordinate systems, and
therefore should not have associated independent central charges in
the Type IIB spacetime supersymmetry algebra.
The same kind of transformation identifies the KK-6A brane with the D6
brane, and the KK-8A brane with the D8-brane. This also explains
why the KK-6A and KK-8A branes
do not appear in the Type IIA spacetime supersymmetry algebra.

The paper is organized as follows. In the first four sections we
present the effective actions of the exotic branes that we have
discussed. We start with the KK-6A brane, which is simply obtained
by reducing the M-theory Kaluza-Klein monopole.
The NS-7B brane is constructed by T-duality from
the KK-6A brane. We see that it is related by S-duality to the D7-brane,
where in this duality transformation the RR 8-form is mapped onto 
another 8-form which is
defined as the T-dual of the gravitational field with respect to
which the KK-6A brane is charged.
We discuss the role played by this field within the Type IIB
theory and discuss some properties of the solutions, in particular
we show that the D7 and NS-7B brane solutions correspond to local versions 
of the weak and strong coupling limits of the non-perturbative solution
of \cite{GGP}. 
In section 4 we construct the effective action of the KK-8A brane and
discuss its interpretation within a non-covariant massive Type IIA 
supergravity.
In section 5 we give an example of an exotic brane with two gauged
isometries which also has a tension $T\sim 1/g_s^3$ and is
related to the other exotic branes by T-duality. This brane is 
predicted by U-duality of M-theory on $T^8$ and should play a role
in the description of eight dimensional massive supergravity. 
For all these branes we
give an analysis of the brane solitons originated by the boundaries
of other branes. 
Finally in section 6 we show that a Kaluza-Klein monopole 
in M-theory with an extra isometric direction can be written 
as a torus bundle over a two-dimensional base
space, and that a modular transformation in the torus can be undone by
a simple coordinate transformation in the base
space. This allows to establish an equivalence between the
branes that can be derived from the M-theory 
Kaluza-Klein monopole after
reduction and T-duality, in particular between the D6 and the
KK-6A branes and the D7 and the NS-7B branes.
We also discuss the relation between the KK-8A brane and the D8-brane,
using the same construction. Finally, we comment on the origin of this
equivalence from the point of view of the M9-brane.


\section{The IIA KK6-brane}
\label{IIA-KK6-brane}


This brane is obtained by reducing the M-theory KK-monopole effective
action along a transverse coordinate.
In the notation of \cite{H1} it corresponds to the $(6,1^2;3)$ 
brane\footnote{In the notation of \cite{H1} 
the first entry gives the number of ordinary spatial directions, 
the entries of the form $m^n$ indicate that there are $m$ spatial
directions which are gauged in the effective action and whose radii
have $nth$ power. Finally the last entry gives the power of the
inverse of the string coupling constant.} predicted by U-duality
of M-theory on $T^7$. In this section we construct its worldvolume
effective action and leave for Section 6 and the Conclusions the
discussion about its role within the Type IIA theory.

The action of the M-KK-monopole was constructed in \cite{BJO,BEL}
and it is given by:
\begin{equation}
\label{MKKaccion}
\begin{array}{rcl}
{\hat S}_{{\rm MKK}} &=&
-{\hat T}_{{\rm MKK}}  \int d^7 \xi \,\,\,
|{\hat k}|^2 \sqrt{|{\rm det}(D_i {\hat X}^{{\hat \mu}} 
D_j {\hat X}^{{\hat \nu}} 
{\hat g}_{{\hat \mu}{\hat \nu}}+ l_p^2
|{\hat k}|^{-1} {\hat {\cal F}_{ij}})|} \\
& & \\
& &
+ \,\, {1  \over 7!}\, l_p^2 \, {\hat T}_{{\rm MKK}} 
\int  d^7 \xi \,\, \epsilon^{i_1 \dots i_7}
\,\, {\hat {\cal K}}^{(7)}_{i_1 \dots i_7} \, , \\
\end{array}
\end{equation}

\noindent with\footnote{$i_{\hat k}{\hat P}$ denotes the interior product
of a field ${\hat P}$ with the Killing vector.}:

\begin{equation}
\label{efe}
\begin{array}{rcl}
{\hat {\cal F}} &=& 2 \partial {\hat \omega}^{(1)} 
+ l_p^{-2}
\partial {\hat X}^{{\hat \mu}} \partial {\hat X}^{{\hat \nu}}
\ikC_{{\hat \mu}{\hat \nu}} \, , \\
\end{array}
\end{equation}

\noindent and:

\begin{equation}
\begin{array}{rcl}
&&{\hat {\cal K}}^{(7)} =
7 \left\{ \partial {\hat \omega}^{(6)} - {1 \over 7}l_p^{-2}
({i}_{\hat k} {\hat N}) 
+3 ({i}_{\hat k}{\hat {\tilde C}}){\hat {\cal F}} \right.
\\ & & \\ 
& &
- 5  l_p^{-2} D{\hat X}^{{\hat \mu}}
D  {\hat X}^{{\hat \nu}}D  {\hat X}^{{\hat \rho}}
{\hat C}_{{\hat \mu} {\hat \nu} {\hat \rho}} \ikC  \ikC 
-30 D {\hat X}^{{\hat \mu}}D {\hat X}^{{\hat \nu}}
D {\hat X}^{{\hat \rho}}
{\hat C}_{{\hat \mu} {\hat \nu} {\hat \rho}} \ikC  \partial 
{\hat \omega}^{(1)}
\\ & & \\ 
& &
\left. -60 l_p^2 D {\hat X}^{{\hat \mu}} D {\hat X}^{{\hat \nu}}
D {\hat X}^{{\hat \rho}} {\hat C}_{{\hat \mu} {\hat \nu}
{\hat \rho}}
\partial {\hat \omega}^{(1)} \partial {\hat \omega}^{(1)} 
-120 l_p^4 
{\hat A} \partial {\hat \omega}^{(1)} \partial 
{\hat \omega}^{(1)} \partial {\hat \omega}^{(1)}
\right\} \, . \\
\end{array}
\end{equation}

\noindent  This action reflects the fact that the Taub-NUT space
of the monopole is isometric in its Taub-NUT direction, so
that the transverse space is effectively three dimensional.
The extra degree of freedom is eliminated by imposing
a Killing isometry in the Taub-NUT direction:

\begin{equation}
\label{iso}
\delta {\hat X}^{\hat \mu}=-{\hat \sigma}^{(0)}
{\hat k}^{\hat \mu}\, ,
\end{equation}

\noindent which is gauged through
the introduction of the dependent gauge field ${\hat A}$
\footnote{In our notation 
$|{\hat k}|^2=-{\hat k}^{\hat \mu}{\hat k}^{\hat \nu}
{\hat g}_{{\hat \mu}{\hat \nu}}$.}:

\begin{equation}
{\hat A}=|{\hat k}|^{-2}\partial {\hat X}^{{\hat \mu}}
{\hat k}_{{\hat \mu}}\, ,
\end{equation} 

\noindent such that
${\hat k}_{\hat \mu}D {\hat X}^{\hat \mu}=0$, with 
$D{\hat X}^{\hat \mu}$ the covariant derivative:

\begin{equation}
D {\hat X}^{{\hat \mu}}=\partial {\hat X}^{{\hat \mu}}
+{\hat A} {\hat k}^{{\hat \mu}}\, .
\end{equation}

\noindent See \cite{BJO,BEL}
for more details.
${\hat C}$ is the 3-form field of eleven dimensional supergravity,
${\hat {\tilde C}}$ its electric-magnetic dual and
${\hat N}$ is the dual of the Killing vector ${\hat k}_{\hat \mu}$
considered as a 1-form. It is therefore a gravitational field.
The presence of covariant derivatives and contractions
with the Killing vector assures invariance under local isometric
transformations: 
$\delta {\hat X}^{\hat \mu}=-{\hat \sigma}^{(0)} (\xi)
{\hat k}^{\hat \mu}$. The field content is then that of the seven
dimensional vector multiplet, involving 3 scalars and 1 vector. 
The gauge transformations of the target space
and worldvolume fields involved in the action can be found in 
\cite{BEL}.

In order to obtain the KK-6A brane effective action we perform a
direct dimensional reduction along a coordinate $y$ different from
the Taub-NUT direction. Accordingly, the transverse space of the 
KK-6A brane still contains 
an isometric Taub-NUT direction, which remains gauged in
the action after the reduction. The field content is again that 
of the seven dimensional vector multiplet, taking into account 
the additional scalar coming from the reduction.

The background fields reduce as follows:
\begin{equation}
\begin{array}{rcl}
(i_{\hat k}{\hat N})_{\mu_1\dots\mu_7}&=&
(i_k N^{(8)})_{\mu_1\dots\mu_7} \\
& & \\
(i_k {\hat N})_{\mu_1\dots\mu_6 y}&=&-(i_k N^{(7)})_{\mu_1\dots\mu_6} \\
& & \\
(i_{\hat k}{\hat {\tilde C}})_{\mu_1\dots\mu_5}&=&
-(i_k B^{(6)})_{\mu_1\dots\mu_5} \\
& & \\
(i_{\hat k}{\hat {\tilde C}})_{\mu_1\dots\mu_4 y}&=&
-(i_k C^{(5)})_{\mu_1\dots\mu_4}+2C^{(3)}_{[\mu_1\mu_2\mu_3}
(i_k B^{(2)})_{\mu_4]}+3(i_k C^{(3)})_{[\mu_1\mu_2}
B^{(2)}_{\mu_3\mu_4]} \\
& & \\
(i_{\hat k}{\hat C})_{\mu_1\mu_2}&=&
(i_k C^{(3)})_{\mu_1\mu_2}\\
& & \\
(i_{\hat k}{\hat C})_{\mu y}&=&-(i_k B^{(2)})_\mu \, . \\
\end{array}
\end{equation}

\noindent $N^{(7)}$ is the electric-magnetic dual of the Killing
vector $k_\mu$, $i_k N^{(7)}$ being the
field to which the IIA KK-monopole couples minimally \cite{BEL}.
Its gauge transformation rule can be found in that reference.
The field $N^{(8)}$ comes from the reduction of ${\hat N}$,
electric-magnetic dual of ${\hat k}_{\hat \mu}$, and must be dual
to the modulus of the Killing vector
(see the Conclusions for a discussion on this point).
Its gauge transformation rule can be found in the Appendix.
Both fields are therefore purely gravitational. $C^{(p)}$ denotes
a $p$-form RR field, $B^{(2)}$ the NS-NS 2-form and $B^{(6)}$ its
electric-magnetic dual. The notation for their gauge transformation
rules is that of \cite{EJL}.

We also have:
\begin{equation}
\begin{array}{rcl}
{\hat {\cal F}}&=&{\cal H}^{(2)} \\
& & \\
{\hat A}&=&A-(2\pi\alpha^\prime)\frac{e^{2\phi}(i_k C^{(1)})}
{|k|^2+e^{2\phi}(i_k C^{(1)})^2}{\cal G}^{(1)} \, , \\
\end{array}
\end{equation}

\noindent where:
\begin{equation}
\label{hachege}
\begin{array}{rcl}
{\cal H}^{(2)}&=&2\partial d^{(1)}+\frac{1}{2\pi\alpha^\prime}
(i_k C^{(3)})-2(i_k B^{(2)})\partial c^{(0)} \\
& & \\
{\cal G}^{(1)}&=&\partial c^{(0)}+\frac{1}{2\pi\alpha^\prime}
(DX C^{(1)}) \, , \\ 
\end{array}
\end{equation}

\noindent and: 
\begin{equation}
c^{(0)}=y/(2\pi\alpha^\prime)\, ,\,\,\,\,
d^{(1)}={\hat \omega}^{(1)} \, .
\end{equation}

The result for the dimensionally reduced action is:
\begin{equation}
\label{m0accion}
\begin{array}{rcl}
S_{{\rm KK6A}}& =&
-T_{{\rm KK6A}} \int d^7 \xi \,\,
e^{-3\phi} |k|^2 (1 + e^{2 \phi}|k|^{-2} (i_k C^{(1)})^2) \times \\
& & \\
& &\hspace{-1.5cm}
\times\sqrt{|{\rm det}(D_iX^\mu D_jX^\nu g_{\mu \nu}
- \frac{(\alfa)^2 e^{2\phi}}{1+e^{2\phi}|k|^{-2}(i_k C^{(1)})^2}
 {\cal G}^{(1)}_i {\cal G}^{(1)}_j +
 { (\alfa) |k|^{-1} e^\phi \over \sqrt{ 1 + e^{2 \phi}|k|^{-2} 
(i_k C^{(1)})^2}}
{\cal H}^{(2)}_{ij})|}\\
& & \\
& & \hspace{-1.5cm}
+ \,\, {1 \over 7!}(\alfa) \, T_{{\rm KK6A}} \, \int d^7\xi
\,\,  \epsilon^{i_1 \dots i_7}
{\cal K}^{(7)}_{i_1 \dots i_7} \, ,\\
\end{array}
\end{equation}

\noindent where $T_{{\rm KK6A}}={\hat T}_{{\rm MKK}}$ 
and ${\cal K}^{(7)}$ is given by:

\begin{equation}
\label{KK6WZ}
\begin{array}{rcl}
{\cal K}^{(7)} &=& 7\left\{ \partial \omega^{(6)} 
-\frac{1}{7(2\pi\alpha^\prime)}(i_k N^{(8)})+
(i_k N^{(7)})\partial c^{(0)}-3(i_k B^{(6)}){\cal H}^{(2)}\right.
\\ & & \\ & &
-15(2\pi\alpha^\prime)(i_k C^{(5)})\partial c^{(0)}{\cal H}^{(2)}
+30(2\pi\alpha^\prime)C^{(3)}(i_k B^{(2)})\partial c^{(0)} {\cal H}^{(2)}
\\ & & \\ & & 
+45(2\pi\alpha^\prime)(i_k C^{(3)})B^{(2)}\partial c^{(0)}{\cal H}^{(2)}-
\frac{5}{2\pi\alpha^\prime}C^{(3)}(i_k C^{(3)})^2
\\ & & \\ & & 
+20 C^{(3)} (i_k C^{(3)})(i_k B^{(2)})\partial c^{(0)}-
30(2\pi\alpha^\prime)C^{(3)}{\cal H}^{(2)}\partial d^{(1)}
\\ & & \\ & & 
-15B^{(2)}\partial c^{(0)} (i_k C^{(3)})^2-90(2\pi\alpha^\prime)^2
B^{(2)}\partial c^{(0)}{\cal H}^{(2)}\partial d^{(1)}
\\ & & \\ & & 
\left. -15(2\pi\alpha^\prime)^2 (A-\frac{(2\pi\alpha^\prime)e^{2\phi}
(i_k C^{(1)})}{|k|^2+e^{2\phi}(i_k C^{(1)})^2}{\cal G}^{(1)})
({\cal H}^{(2)})^3 \right\} \, .\\
\end{array}
\end{equation}

\noindent The covariant derivative is defined as:

\begin{equation}
D X^\mu=\partial X^\mu+A k^\mu
\end{equation}

\noindent with $A$ given by:

\begin{equation}
A=|k|^{-2}\partial X^\mu k_\mu\, .
\end{equation}
Finally, the effective tension of this brane is given by:
\be
{\cal T}_{KK6A} = {R_k^2 \over (2\pi)^6 g_s^3 l_s^9} \, ,
\ee 
where $R_k$ is the radius of the compact direction associated to 
the Taub-NUT isometry and $l_s$ is the string length.

We find that the worldvolume field content is that of the seven
dimensional vector multiplet, namely
a scalar $c^{(0)}$, a vector $d^{(1)}$ and a 
6-form $\omega^{(6)}$ playing the role of tension of the brane.
These fields can be given an interpretation in terms of solitons
in the brane. In general, worldvolume fields coupled to a 
given brane effective action describe soliton solutions in its
worldvolume. For branes with a Killing
direction, the worldvolume fields enter in the field strengths
as:

\begin{equation}
\label{fieldstre}
{\cal K}^{(p)}=p\partial\omega^{(p-1)}+\frac{1}{2\pi\alpha^\prime}
(i_k C^{(p+1)})+\dots \, ,
\end{equation}

\noindent from where we can conclude that $\omega^{(p-1)}$ couples to a 
$(p-2)$-brane soliton describing the boundary of a $p$-brane
ending on the monopole, with one of its worldvolume directions
wrapped around the Killing direction \cite{EJL}.
Therefore $c^{(0)}$ must
couple to a 4-brane soliton (through its worldvolume dual),
$d^{(1)}$ to a 0-brane and a 3-brane solitons, and $\omega^{(6)}$
to a domain-wall  soliton. The corresponding
configurations of branes ending on the KK-6A brane
can be associated to some of the intersections of the KK-6A
brane with other Type IIA branes. These were classified in
\cite{Groningen-Boys-2}. We find:

\begin{itemize}

\item The intersections $(0|{\rm D2},{\rm KK6A})$ 
and $(0|{\rm F1},{\rm KK6A})$ are associated to a
0-brane soliton coupled to $d^{(1)}$. 
These configurations are obtained from the
intersection $(0|{\rm M2},{\rm MKK})$ in M-theory.

\item $(3|{\rm D4},{\rm KK6A})$ and $(3|{\rm NS5A},{\rm KK6A})$
are associated to a 3-brane soliton coupled magnetically to
$d^{(1)}$. They are obtained from the configuration $(3|{\rm M5},{\rm MKK})$.

\item $(5|{\rm NS5A},{\rm KK6A})$ 
represents a NS-5A brane embedded in the KK-6A and
it is associated to a domain-wall
soliton which couples to $\omega^{(6)}$. 
It is obtained from the configuration $(5|{\rm M5},{\rm MKK})$.

\item Further, we find three configurations that can give rise to
a 4-brane soliton. The cases $(4|{\rm KK6A},{\rm KK6A})^a$ 
and $(4|{\rm KK5A},{\rm KK6A})^b$ couple magnetically to the embedding scalars
and $(4|{\rm D6},{\rm KK6A})$ couples magnetically to the form $c^{(0)}$.
The index $a$ ($b$) indicates that the isometries of both
objects lie in the same (a different) direction. 
These configurations are obtained from the configurations 
$(4|{\rm MKK},{\rm MKK})^{a,b}$ in M-theory
through reduction along a transverse direction \cite{Groningen-Boys-2}.

\end{itemize}


\section{The IIB NS-7 brane}
\label{IIB-NS-7-brane}


As we have mentioned in the Introduction, the NS-7B brane 
can be obtained
by T-duality from the KK-6A brane. The T-duality transformation
takes place along the Taub-NUT direction of the KK-6A brane and
the result is the double dimensional reduction of the NS-7B
brane.
This brane plays a role in the construction of the massive
nine dimensional Type II supergravity of \cite{PT}, which is
manifestly invariant under $SL(2,\R)$. This theory contains three
mass parameters, only two of which are independent upon
quantization, and they have their origin in the presence of D7 and
NS-7B branes in the background.
Also, as we discuss in the end of this section,
the orientifold constructions of the Type IIB
theory on $T^2$ describing certain branches of F-theory on K3
require the introduction of NS-7B spacetime filling
branes to cancel the charge of the orientifold fixed planes
at strong coupling.


\subsection{The NS-7B Brane Effective Action}


In this section
we construct the worldvolume effective action of the NS-7B brane
by performing a T-duality on the KK-6A brane effective action.
In the next subsections we show 
that it is connected to the D7-brane by
an S-duality transformation and discuss in some extent the role
that this brane plays within the Type IIB theory.

The result of a T-duality transformation in the effective action
of the KK-6A brane along its Taub-NUT direction
is the (double dimensional reduction of the) action:

\begin{equation}
\label{NS7B}
\begin{array}{rcl}
S_{{\rm NS7}} & =&
-T_{{\rm NS7}} \int d^8 \xi \,\,
e^{-3\phi} (1 + e^{2 \phi}(C^{(0)})^2)
\sqrt{|{\rm det}(g+\frac{(\alfa) e^{\phi}}
{\sqrt{1+e^{2\phi}(C^{(0)})^2}}{\tilde {\cal F}})|}\\
& & \\
& & \hspace{-1.5cm}
+ \,\, {1 \over 8!}(\alfa) \, T_{{\rm NS7}} \, \int d^8\xi
\,\,  \epsilon^{i_1 \dots i_8}
{\tilde {\cal G}}^{(8)}_{i_1 \dots i_8} \, ,\\
\end{array}
\end{equation}

\noindent where $T_{{\rm NS7}}$ is such that 
$T_{{\rm NS7}}\int d\xi^8=T_{{\rm KK6A}}$ and
${\tilde {\cal G}}^{(8)}$ is given by:

\begin{equation}
\label{Gtilde8}
\begin{array}{rcl}
&&{\tilde {\cal G}}^{(8)} = \left\{
8\partial {\tilde c}^{(7)} - {1 \over \alfa}{\tilde C}^{(8)}
+\frac12\frac{8!}{6!} B^{(6)} {\tilde {\cal F}}+\frac{7!}{4!}
(2\pi\alpha^\prime)C^{(4)}{\tilde {\cal F}}^2 
\right. \\ & &\\
& & +\frac12 \frac{7!}{3!}(2\pi\alpha^\prime)^2 B^{(2)}
{\tilde {\cal F}}^3-\frac12 \frac{7!}{4!}(2\pi\alpha^\prime)^3
\frac{C^{(0)}}{(C^{(0)})^2+e^{-2\phi}}{\tilde {\cal F}}^4
-\frac{7!}{8}(2\pi\alpha^\prime)B^{(2)}C^{(2)}{\tilde {\cal F}}^2+
\\& &\\& &
\left.
+\frac{7!}{8}B^{(2)}(C^{(2)})^2{\tilde {\cal F}}
-\frac12 \frac{7!}{3!}\frac{1}{2\pi\alpha^\prime}B^{(2)}(C^{(2)})^3
\right\} \, .\\
\end{array}
\end{equation}

The T-duality transformation rules of the background fields 
that couple to the KK-6A brane can be
found for instance in \cite{EJL}, whereas the transformation rule of the
gravitational field $N^{(8)}$ has to be worked out, 
with the result that it transforms as:

\begin{equation}
\label{eneocho}
\begin{array}{rcl}
(i_k N^{(8)})_{\mu_1\dots\mu_7}^\prime&=&
{\tilde C}^{(8)}_{\mu_1\dots\mu_7 z}-7(B^{(6)}_{[\mu_1\dots\mu_6}
-6 B^{(6)}_{[\mu_1\dots\mu_5 z}\frac{g_{\mu_6 z}}{g_{zz}})
C^{(2)}_{\mu_7] z}\\
& & \\
& & +35C^{(4)}_{[\mu_1\dots\mu_3 z}C^{(2)}_{\mu_4\mu_5}
(C^{(2)}_{\mu_6\mu_7]}+4\frac{g_{\mu_6 z}}{g_{zz}}C^{(2)}_{\mu_7] z})\\
& & \\
& &-105(2B^{(2)}_{[\mu_1\mu_2}-B^{(2)}_{[\mu_1 z}\frac{g_{\mu_2 z}}{g_{zz}})
C^{(2)}_{\mu_3 z}(C^{(2)})^2_{\mu_4\dots\mu_7]}\, .\\
\end{array}
\end{equation}

\noindent This defines a new field ${\tilde C}^{(8)}$ in Type IIB,
which as we will discuss below is the S-dual of the RR 8-form.

The worldvolume fields transform as follows:

\begin{equation}
c^{(0)\prime}=-c^{(1)}_\sigma \, , \,\,\,\, d^{(1)\prime}=-c^{(1)}
\end{equation}

\noindent so that:

\begin{equation}
{\cal G}_i^{(1)\prime}=-{\tilde {\cal F}}_{i\sigma}\, , \,\,\,
{\cal H}^{(2)\prime}_{ij}=-{\tilde {\cal F}}_{ij}-
2\frac{g_{[iz}}{g_{zz}}{\tilde {\cal F}}_{j]\sigma}\, ,
\end{equation}

\noindent with:

\begin{equation}
{\tilde {\cal F}}=2\partial c^{(1)}+\frac{1}{2\pi\alpha^\prime}C^{(2)}
\end{equation}

\noindent the field strength associated to the RR 2-form. The tensions
transform simply as:

\begin{equation}
\omega^{(6)\prime}_{i_1\dots i_6}={\tilde c}^{(7)}_{i_1\dots i_6\sigma}\, .
\end{equation}

{}From (\ref{Gtilde8}) we see that the NS-7B brane is minimally
coupled to the field ${\tilde C}^{(8)}$, arising
from the T-duality transformation (\ref{eneocho}). The interpretation
of this new field in the IIB theory is discussed in the next
subsection. (\ref{NS7B}) and (\ref{Gtilde8}) also identify 
the strings that can end on the
brane as D-strings, so that this brane corresponds to the 
(0,1) 7-brane in the notation of \cite{DL}.

Before comparing with the D7-brane effective action let us
analyze the possible soliton solutions that can occur in the
worldvolume of this brane. Its worldvolume field content consists
on $c^{(1)}$ and ${\tilde c}^{(7)}$,
playing the role of tension
of the brane, as well as the embedding scalars. They couple to
the soliton configurations:

\begin{itemize}

\item $(0|{\rm D1},{\rm NS7B})$ gives a 0-brane soliton coupled to $c^{(1)}$.
It is related to the configuration $(0|{\rm D2},{\rm KK6A})$ by T-duality.

\item $(4|{\rm D5},{\rm NS7B})$ gives a 4-brane soliton coupled 
magnetically to $c^{(1)}$. It is related by T-duality to
$(4|{\rm D6},{\rm KK6A})$.

\item $(5|{\rm NS7B},{\rm NS7B})$ and $(5|{\rm KK5B},{\rm NS7B})$
give a 5-brane soliton magnetically coupled to
the embedding scalars. They are related by T-duality to
$(4|{\rm KK6A},{\rm KK6A})^a$, and $(4|{\rm KK5A},{\rm KK6A})^b$,
$(5|{\rm NS5A},{\rm KK6A})$ respectively.

\item Note that the presence of ${\tilde c}^{(7)}$ as a dynamical tension is
not entirely justified. The existence of such a dynamical tension implies 
that the NS-7B brane can be open \cite{Townsend}, hence the NS-7B
(and by S-duality, the D7-brane) 
would have boundaries. However,
this would not define consistently the monodromies associated to these
7-branes. 

\end{itemize}


\subsection{D7-branes, NS7-branes and S-duality}
\label{sduals}


The effective action that we have found for the NS-7B brane is related
by S-duality to the effective action of the D7-brane.

\begin{equation}
S_{{\rm D7}}=-T_{D7}\int d^8 \xi\ \Bigl\{ e^{-\phi}\sqrt{|{\rm
    det}(g+(2\pi\alpha^\prime){\cal F})|}+
\frac{(2\pi\alpha^\prime)}{8!}\epsilon^{i_1\dots i_8}
{\cal G}^{(8)}_{i_1\dots i_8}\Bigr\}\, ,
\end{equation}

\noindent with 
${\cal F}=2\partial b+\frac{1}{2\pi\alpha^\prime}B^{(2)}$ and
${\cal G}^{(8)}$ given by:
\begin{equation}
\label{WZD7}
\begin{array}{rcl}
{{\cal G}}^{(8)} &=& \left\{
8\partial {c}^{(7)} + {1 \over \alfa}C^{(8)}
-\frac12 \frac{8!}{6!}C^{(6)}{\cal F}+\frac{7!}{4!}
(2\pi\alpha^\prime)C^{(4)}{\cal F}^2 
\right. 
\\ & &\\& & 
-\frac12 \frac{7!}{3!}(2\pi\alpha^\prime)^2 C^{(2)}{\cal F}^3
+\frac12 \frac{7!}{4!}(2\pi\alpha^\prime)^3 C^{(0)}{\cal F}^4
\\& &\\& &
\left. +\frac{7!}{8}(2\pi\alpha^\prime)C^{(2)}B^{(2)}{\cal F}^2
-\frac{7!}{8}C^{(2)} (B^{(2)})^2 {\cal F}+\frac12 \frac{7!}{3!}
\frac{1}{2\pi\alpha^\prime}(B^{(2)})^3 C^{(2)}
\right\} \, ,\\
\end{array}
\end{equation}

\noindent yields (\ref{NS7B}) 
after an S-duality transformation:
\begin{equation}
\label{Strans}
\begin{array}{rcl}
C^{(0)} &\rightarrow& {- C^{(0)} \over (C^{(0)})^2 + e^{-2 \varphi}} \, ,\\
& &\\
C^{(2)} &\rightarrow& B^{(2)} \, ,\\
& &\\
C^{(6)} &\rightarrow& B^{(6)}\, ,\\
\end{array}
\hspace{1.5cm}
\begin{array}{rcl}
e^{-\varphi} &\rightarrow& { e^{- \varphi} \over (C^{(0)})^2 + e^{-2 \varphi}}
\, ,\\
& &\\
B^{(2)} &\rightarrow& - C^{(2)} \, ,\\
& &\\
B^{(6)} &\rightarrow& - C^{(6)} \, ,\\
\end{array}
\end{equation}
\begin{displaymath}
\begin{array}{rcl}
C^{(4)} \rightarrow C^{(4)}\, ,\\
& &\\
C^{(8)} \rightarrow -{\tilde C}^{(8)}\, ,\\
& &\\
{\tilde C}^{(8)} \rightarrow -C^{(8)}\, ,\\
\end{array}
\end{displaymath}

\noindent together with:
\begin{equation}
\begin{array}{rcl}
b &\rightarrow&  -c^{(1)}  \, ,\\
& &\\
c^{(1)} &\rightarrow& b \, ,\\
\end{array}
\hspace{2.5cm}
\begin{array}{rcl}
c^{(7)} &\rightarrow& {\tilde c}^{(7)} \, ,\\
& &\\
{\tilde c}^{(7)} &\rightarrow& c^{(7)} \, ,\\
\end{array}
\end{equation}

\noindent for the worldvolume fields.

One can check as well that the soliton configurations 
that we have found for the NS-7B brane
are mapped under S-duality to the usual 
intersections involving the D7-brane.

The transformation rules for $C^{(8)}$ and ${\tilde C}^{(8)}$
in (\ref{Strans}) show that the D7 and NS-7B branes do not 
form a doublet under S-duality. They imply in particular 
that after two S-duality transformations each WZ term is 
mapped onto itself,
whereas for doublets $S^2=-1$.
This observation from the worldvolume effective actions is in
agreement with the results of \cite{PT}, where it is shown that
Type IIB 7-branes transform as triplets under SL(2,\Z).
We now elaborate on this point.

Consider the conserved currents (in Einstein frame)
associated to the RR scalar 
and dilaton of the Type IIB theory:

\begin{eqnarray}
\label{corrientes}
&&j^{(0)}=e^{2\phi}d C^{(0)} \nonumber\\
&&j^{(\phi)}=e^{2\phi} d |\lambda|^2\, .
\end{eqnarray}

\noindent Here $\lambda=C^{(0)}+ie^{-\phi}$ is the axion-dilaton
modulus.
These currents define the electric-magnetic duals of the
RR scalar and dilaton, namely the fields
$C^{(8)}$, to which the D7-brane couples minimally, and a certain
8-form $B^{(8)}$, which in principle could be associated to another
7-brane in the theory.
There is a third conserved current \cite{PT} which can be obtained
as a combination of the previous two with non-constant coefficients:

\begin{equation}
\label{tercera}
j =-C^{(0)}j^{(\phi)}+|\lambda|^2 j^{(0)}\, ,
\end{equation}

\noindent to which we can associate as well a dual field. 
The transformation rules of these currents under 
the S and T generators of
the $SL(2,\Z)$ group are \cite{PT}:

\begin{equation}
\begin{array}{rcl}
&&\\
&&\\
&{\rm S}:& \\
&&\\
&&\\
\end{array}
\begin{array}{rcl}
j^{(0)} &\rightarrow& -j \, \\
& &\\
j^{(\phi)} &\rightarrow& -j^{(\phi)} \, \\
& &\\
j &\rightarrow& -j^{(0)} \, \\
\end{array}
\begin{array}{rcl}
&&\\
&&\\
&{\rm T}:& \\
&&\\
&&\\
\end{array}
\begin{array}{rcl}
\label{Stransforms}
j^{(0)} &\rightarrow& j^{(0)} \, \\
& &\\
j^{(\phi)} &\rightarrow& j^{(\phi)} + 2 j^{(0)}\, \\
& &\\
j &\rightarrow& j - j^{(0)} - j^{(\phi)}\, \\
\end{array}
\end{equation}

\noindent The transformations of $j^{(0)}$ and $j$ under S suggest
that the 8-form associated to $j$ should be the S-transformed
of $C^{(8)}$, namely ${\tilde C}^{(8)}$. Indeed, 
with this identification 
the S-duality transformations of the 8-forms become:

\begin{equation}
\ba{rcl}
&&\\
&&\\
&{\rm S}:& \\
&&\\
&&\\
\ea
\begin{array}{rcl}
\label{ochofields1}
C^{(8)} &\longrightarrow& -{\tilde C}^{(8)}
\\& &\\
{\tilde C}^{(8)} &\longrightarrow& -C^{(8)} 
\\& &\\
B^{(8)} &\longrightarrow& -B^{(8)} \, ,
\ea
\ee

\noindent which agree with the transformation rules for $C^{(8)}$
and ${\tilde C}^{(8)}$ in (\ref{Strans}).
We see that a 7-brane
magnetically coupled to the dilaton would transform as a singlet
under S-duality, and that the 8-forms $C^{(8)}$ and ${\tilde C}^{(8)}$
do not transform as a doublet.
In fact, the 8-forms transform as a triplet under S-duality,
and in general, under the $SL(2,\Z)$ symmetry group:

\begin{equation}
{\cal M}=\left( \begin{array}{cc}
B^{(8)} &  2{\tilde C}^{(8)} \\
2C^{(8)} & -B^{(8)} \end{array} \right)
\rightarrow \Lambda {\cal M} \Lambda^{-1}
\, , \qquad \forall \Lambda \in SL(2,\Z) \, .
\end{equation}

The field ${\tilde C}^{(8)}$ becomes necessary when we try to map 
$C^{(8)}$ onto a local field under S-duality, but
of course the three 8-forms are not independent, since they are
related through (\ref{tercera}), which in terms of 8-forms reads:

\begin{equation}
\label{nolocal}
d{\tilde C}^{(8)}=-C^{(0)} dB^{(8)}+|\lambda|^2 dC^{(8)}\, .
\end{equation}

\noindent Therefore ${\tilde C}^{(8)}$ is not interpreted as an
independent field in the theory.
A dual Type IIB supergravity, written as
a function of the dual potentials, will depend only on two of the three
9-form field strengths\footnote{As explained in 
\cite{D-Lechner-Tonin} this
formulation will depend as well on the original fields through
the Chern-Simons forms in the definition of the curvatures.},
which is consistent with the results of \cite{CGNSW} and 
\cite{D-Lechner-Tonin}\footnote{In this last reference three 
different 9-form field strengths appear in the formulation of 
Type IIB supergravity in terms of original and dual fields due to
the fact that the action is manifestly SU(1,1) invariant with the
U(1) invariance not being fixed.}.
This relation only becomes local for constant axion-dilaton modulus:

\begin{equation}
{\tilde C}^{(8)}=-C^{(0)}B^{(8)}+|\lambda|^2 C^{(8)}\, .
\end{equation}

In order to clarify further the role of this three 8-form potentials
let us consider the solution
representing a single D7-brane located at the origin of the
transverse space \cite{PT,Vafa} (in string frame):

\begin{equation}
\label{solucion}
\begin{array}{rcl}
ds^2 &=& H^{-1/2}ds_{1,7}^2 -H^{1/2}d\omega d{\bar \omega}  \, ,\\
& &\\
\lambda &=& \frac{1}{2\pi i}\log{\omega}\, ,\\
\end{array}
\end{equation}

\noindent where the harmonic function is given by 
$H=-\frac{1}{2\pi}\log{|\omega|}$ with 
$\omega=x_2+ix_1$, and $x_1, x_2$ the two transverse coordinates.
The RR charge of this brane can be computed as:

\begin{equation}
p=\oint_\gamma \,^*dC^{(8)}
\end{equation}

\noindent where $\gamma$ is a circle around the origin in the
transverse space.
A 7-brane with a non-vanishing value for
the dilaton carries as well magnetic dilaton charge, which is
computed as:

\begin{equation}
\label{cargas}
s=\oint_\gamma \,^*dB^{(8)}\, .
\end{equation}

\noindent Moreover,
through (\ref{nolocal}) it also carries ${\tilde C}^{(8)}$-charge:

\begin{equation}
q=\oint_\gamma \,^*d{\tilde C}^{(8)}=
\oint_\gamma (-C^{(0)}\,^*dB^{(8)}+|\lambda|^2\,^*dC^{(8)}).
\end{equation}

\noindent The values of the three charges can be computed with the
only input of the axion-dilaton modulus associated to the solution. 
The quantization of the charges is
obtained by imposing that $e^Q$ with 
\begin{equation}
Q=\left( \begin{array}{cc}
s &  2p \\
2q & -s \end{array} \right)
\end{equation}

\noindent belongs to $SL(2,\Z)$. Imposing this condition it is
possible to see that
the three charges can be obtained from only two independent integers
(see Section 5.1 in \cite{PT}) whose values specify uniquely the
7-brane charges. It also implies 
that it is not possible to have a solution with vanishing
$p$ and $q$ charges and non-vanishing $s$,
which is consistent with the fact that a 7-brane carrying only magnetic
dilaton charge cannot exist, given that
(\ref{cargas}) vanishes for $C^{(0)}=0$. 
Taking into account that 7-branes with the 
same monodromy matrix are equivalent,
the charge matrix is further restricted to have the
form \cite{DL}:

\begin{equation}
\Lambda\left( \begin{array}{cc}
1 &  1 \\
0 & 1 \end{array} \right)\Lambda^{-1}
\, , \qquad \forall \Lambda \in SL(2,\Z) \, .
\end{equation}

If we compute the values of the charges carried by
the D7-brane solution above:

\begin{equation}
\begin{array}{rcl}
p &=&\oint_\gamma e^{2\phi}dC^{(0)}\, ,\\
& & \\
q &=&\oint_\gamma e^{2\phi}(-C^{(0)}d|\lambda|^2
+|\lambda|^2dC^{(0)}) \, ,\\
& & \\
s &=&\oint_\gamma e^{2\phi}d|\lambda|^2\, ,\\
\end{array}
\end{equation}

\noindent we find that $p=s=4\pi^2(\log{r})^{-2}$
and $q=1-p/3$,
where we have parametrized $\omega=re^{i\theta}$. 
Therefore asymptotically
$p$ and $s$ vanish whereas the $q$ charge tends to a constant 
value\footnote{In string frame $p=s=1$ and 
$q=-\frac13 +\frac{1}{g_{st}^2}$. However the Einstein frame is more
useful in order to study the S-duality properties of the solution.}.
This is consistent with the fact that asymptotically the
string coupling constant vanishes so that we are in a weak coupling
regime, in which ${\tilde C}^{(8)}$ is not well-defined.
In the strong coupling limit the solution that makes sense is the S-dual
of the D7-brane solution above, which has
the following form \cite{PT}:

\begin{equation}
\label{Ssolucion}
\begin{array}{rcl}
d{\tilde s}^2 &=& (H^2+\theta^2/(2\pi)^2)^{1/2}
[H^{-1/2}ds_{1,7}^2-H^{1/2}d\omega d{\bar \omega}] \, ,\\
& &\\
{\tilde \lambda} &=& -2\pi i / \log{\omega} \, .\\
\end{array}
\end{equation}

\noindent Here the harmonic function is given by
$H=-\frac{1}{2\pi}\log{|\omega|}$, as before.
This solution carries as well $({\tilde p},{\tilde q},{\tilde s})$ 
charges, which in the Einstein
frame take the values: ${\tilde q}={\tilde s}=-4\pi^2(\log{r})^{-2}$,
${\tilde p}=-(1+{\tilde q}/3)$, i.e. they are related to the
charges of the D7-brane solution by the S-duality transformation in
(\ref{Stransforms}). In this case it is the ${\tilde p}$ charge
the one that has a constant value at infinity, whereas 
${\tilde q}={\tilde s}=0$.

Let us now clarify the relation between the D7 and NS-7B brane
solutions that we have just considered and the finite energy
7-brane solution of \cite{GGP}\footnote{We will take 
however the conventions in \cite{Dab}.}.
In order to construct a finite energy 7-brane solution,
where the modulus $\lambda$ depends holomorphically on the transverse space, 
it is crucial to allow $\lambda$ to have an ambiguity
under $SL(2,\Z)$ transformations \cite{GGP,Greene-etal}.
Then the solution is constructed in such a way that the
axion-dilaton modulus parametrizes
the fundamental region of $SL(2,\Z)$, therefore all choices related
by an $SL(2,\Z)$ transformation have been identified, and there is
a one to one mapping from the fundamental region of $SL(2,\Z)$ and
the complex plane. This is 
achieved by fulfilling the condition 
${\bar \partial}\lambda (\omega,{\bar \omega})=0$, derived
from the equations of motion, implicitly through
the elliptic modular function:

\begin{equation}
\label{jomega}
j(\lambda(\omega))=\frac{1}{\omega}
\end{equation}

\noindent with $j$ given by:

\begin{equation}
j(\tau)=
\frac{\left(\theta_2(\tau)^8+\theta_3(\tau)^8+\theta_4(\tau)^8\right)^3}
{\eta(\tau)^{24}}\, ,
\label{j-function}
\end{equation}

\noindent instead of simply taking $\lambda$ as a function of $\omega$.
$\theta_i$ are the Jacobi $\theta$-functions and
$\eta$ is the Dedekind $\eta$ function:

\begin{equation}
\eta(\tau)=q^{1/24} \prod_{n=1}^{\infty} 
(1-q^n)\, , \qquad q=e^{2\pi i \tau}\, .
\end{equation}

\noindent The metric associated to this solution has a deficit angle
of $\pi/6$ near $\omega=0$, so that taking precisely 
24 7-branes the plane is curled
up into a 2-sphere and one ends up with a compact manifold. 
This compactification can equivalently be described as 
an elliptically fibered K3, with $S^2$ the base of the fiber and
$\lambda$ the $\tau$ parameter of the fiber torus.
This defines F-theory compactified on K3,
and the 24 7-branes are commonly referred to as F-theory
7-branes \cite{Vafa}.

Close to $\omega=0$ $j$ has a pole, which must correspond to
$q=0$, i.e. $Im\lambda \sim \infty$, in the left hand side
of (\ref{jomega}). Near $q=0$ $j(\lambda)$ takes the form:

\begin{equation}
j(\lambda)\sim e^{-2\pi i\lambda} \, ,
\end{equation}

\noindent so that we can determine the precise
form of $\lambda(\omega)$:

\begin{equation}
\lambda=\frac{1}{2\pi i}\log{\omega}\, .
\label{axion-dilaton-D7-brane}
\end{equation}

\noindent This is the axion-dilaton modulus of the D7-brane solution
(\ref{solucion}).

In order to study the form of the solution at strong coupling we use that
$j(-1/\tau) = j(\tau)$ and consider the limit $|\tau| \sim 0$, 
so that close to $\omega=0$ $j(\lambda)$ behaves as:

\begin{equation}
j(\lambda)\sim e^{2\pi i/\lambda}\, ,
\end{equation}

\noindent for $\lambda\rightarrow 0$. 
Therefore near $\omega=0$ we have:

\begin{equation}
\lambda=-\frac{2\pi i}{\log{\omega}}\, ,
\label{axion-dilaton-NS7B-brane}
\end{equation}

\noindent which corresponds to the NS-7B brane solution 
(\ref{Ssolucion}). 
Therefore the D7 and NS-7B brane solutions are just 
the weak and strong coupling limits of the non-perturbative
7-brane solution of \cite{GGP}. We will give further evidence 
in Section 6 for the fact
that they represent the same physical object but in different
coordinates patches, where changing the patch amounts to choosing a 
different region to parametrize the $SL(2,\Z)$ moduli space.

{}From the point of view of the low energy effective actions
we can conclude that the introduction of 
three 8-form potentials is necessary when we want to study
the transformation of the D7-brane under $SL(2,\Z)$, in particular
when trying to map the $C^{(8)}$ field onto a local potential. 
The NS-7B brane effective action that we have constructed should be
interpreted as the source for a 7-brane at strong coupling, where
the $C^{(8)}$ field becomes ${\tilde C}^{(8)}$, which is non-local
in terms of $C^{(8)}$ and $B^{(8)}$.
Moreover, although $B^{(8)}$ has a meaning as the electric-magnetic
dual of the dilaton, ${\tilde C}^{(8)}$ cannot be considered as
an independent field in the Type IIB theory (contributing with one
unit to the counting of the bosonic
degrees of freedom) given that it is a function of the other
8-forms. This gives already an indication that the NS-7B brane 
should not have associated a
central charge in the Type IIB spacetime supersymmetry algebra.
We give further evidence to this suggestion in section 
\ref{MKK-domain-wall}, by showing that the two solutions are related
by a coordinate transformation in the transverse space which is
equivalent to changing the 
fundamental region of $SL(2,\Z)$ by its S-dual.
Similar observations will also apply for the KK-6A and KK-8A branes
related to the NS-7B brane by T-duality.


\subsection{NS-7B Branes and Orientifold Constructions}
\label{orienti}



There are particular limits of F-theory on K3 for which  
$\lambda$ remains constant over the base.  
In these limits  K3 degenerates
into a certain orbifold of $T^4$, which can be made equivalent
to an orientifold of Type IIB on $T^2$. 
The most clear case is that of
F-theory on $T^4/\Z_2$ \cite{Sen}. Sen showed that in this
branch F-theory is equivalent to
Type IIB on $T^2$ modded out by the symmetry 
$(-1)^{F^s_L}\Omega I_{89}$, where $F^s_L$ is the spacetime fermion number
of the left-movers, $\Omega$ is the worldsheet parity reversal
and $I_{89}$ is the spacetime inversion in the two directions
of the torus. In this Type IIB orientifold there are 4 fixed-planes
whose charge is cancelled by the addition of 4 
D7-branes (plus their mirrors). 
Open strings ending on these 7-branes provide $SO(8)^4$
Chan-Paton factors, and this gauge structure coincides with the 
gauge symmetry of
F-theory on $T^4/\Z_2$. Therefore in this particular branch it is
possible to give a description of the gauge structure of the theory
in terms of BPS states in weakly coupled Type IIB string theory.

The situation is not so simple in other branches, where the symmetry
by which one has to mod out in the Type IIB description involves 
non-perturbative operations \cite{DM}. This is what happens in F-theory
on $T^4/\Z_4$, realized as an orientifold of Type IIB on $T^2$ modded by
the $\Z_4$ group generated by  
$SR^{(4)}$, where $S$ is the S-duality generator and $R^{(4)}$ rotates
$\pi/2$ in the 8,9 plane. Another case is F-theory on $T^4/\Z_6$, 
which corresponds to Type IIB on $T^2/\Z_6$, where
the group is generated by $(ST)R^{(6)}$, where $T$ is the T-generator 
of $SL(2,\Z)$ and
$R^{(6)}$ rotates $\pi/3$ in the 8,9 plane. In these limits of
F-theory the orientifold fixed planes in the
Type IIB description are non-perturbative and they require 
the addition of non-perturbative 7-branes.
The analysis of the singularities on the F-theory side reveals
that on the $T^4/\Z_4$ branch the gauge structure is
$E_7\times E_7\times SO(8)$, and in $T^4/\Z_6$, 
$E_8\times E_6\times SO(8)$.

The Type IIB BPS states responsible for the gauge symmetries of
these branches have been identified in \cite{Jo}, as well as the
way the $A_n$ gauge groups, associated to $n$ coinciding branes,
are enhanced so as to recover the gauge group associated to the
singularity structure (see also \cite{Gaberdiel-Zwiebach}).
These BPS states correspond to three different types of 
non-perturbative 7-branes, denoted as $A, B, C$,
such that for a $D_n$ singularity there are $A^nBC$ 7-branes
located at the singularity, and $A^{n-1}BC^2$ for a $E_n$
singularity. At weak coupling the $BC$ 7-branes collapse to
an orientifold plane O \cite{Driscoll} and the gauge
group that emerges is the full $SO(2n)$ for a $D_n$ singularity
and a subgroup $SO(2(n-1))\times U(1)$ for a $E_n$, which can be
associated to Chan-Paton factors of fundamental strings, given that
the $A^n$ branes are D7-branes, located on top of the O fixed plane.
Therefore for $D_n$ singularities the gauge structure is completely
captured perturbatively.
In the strong coupling limit the same reasoning applies, but now
the $A^n$ branes are NS-7B branes, located on top of the
S-dual orientifold fixed planes, and the Chan-Paton
factors are associated to open D-strings.
At intermediate couplings the gauge groups are enhanced,
essentially because it
is necessary to consider different sheets to describe the 7-branes,
and then the same 7-brane can be seen as a different brane in
another sheet, so that there can be open strings connecting them
through complicated paths \cite{Jo} (see \cite{Gaberdiel-Zwiebach}
for a description in terms of string junctions).



\section{The IIA KK8-brane}
\label{IIA-KK-8-brane}


The NS-7B brane 
predicts the existence of a new brane in  Type IIA
which can be obtained from it by T-duality. This brane
is also related to the M9-brane as
follows \cite{H1}. The effective action
of a single M9-brane contains a gauged direction in its
worldvolume \cite{H1,BvdS} in such a way that the
field content is that of the nine dimensional vector multiplet.
Reduction along this gauged direction gives the D8-brane effective
action, whereas the reduction along the transverse direction gives the
action of the NS-9A brane. There is a third possibility,
namely the reduction along a worldvolume direction other than
the Killing direction. This gives an 8-brane with a 
gauged direction in its worldvolume, inherited from the M9-brane,  
which we denote as a KK-8A brane.
In the notation of \cite{H1} it corresponds to the $(7,1^3;3)$
brane predicted by U-duality of M-theory on $T^8$.
The presence of this kind of 8-branes in the background is the
origin of the mass of the massive Type IIA supergravity that is
obtained by reducing the massive eleven dimensional supergravity
of \cite{BLO} along a direction different from the Killing
direction. 

As explained in \cite{BLO} it is necessary to introduce
a Killing direction in order to formulate a massive supergravity in
eleven dimensions. The theory is therefore not fully covariant and 
this is the way the no-go theorem of \cite{BDHS} is circumvented.
The origin of the mass is the presence of M9-branes in the background,
which couple as well to the Killing vector in their worldvolumes.
Romans massive supergravity \cite{Romans} is obtained by reducing 
along the Killing direction, therefore is fully covariant, with the
presence of D8-branes in the background being the origin of the mass
\cite{BdRGPT}. On the other hand, the reduction along a direction
different from the Killing direction gives rise to a non-covariant
Type IIA supergravity, containing a Killing vector. The presence of
KK-8A branes in the background is the origin of the mass of this
theory. 

Massive Type II nine dimensional supergravity is obtained
by reducing Romans massive supergravity to nine dimensions, being
the presence of D7-branes in the background the origin of its mass
\cite{BdRGPT}. If instead
one reduces the massive Type IIA supergravity with Killing direction 
and KK-8A branes along the Killing direction,
one gets a massive 9 dim supergravity in which the mass is
induced by the presence of NS-7B branes,
T-dual to the KK-8A branes.
This theory is S-dual to the massive nine dimensional supergravity
coming from the reduction of Romans massive IIA supergravity,
given that from the eleven dimensional
point of view one is just interchanging the two compactified 
directions. In order to
obtain a manifestly $SL(2,\R)$ invariant massive 9 dim supergravity, 
in which both D7 and NS-7B branes are present in the background, it is
necessary to perform a generalized dimensional reduction, as
explained in \cite{PT}. 
 
One way to construct the action of the KK-8A brane is by
dimensional reduction from the M9-brane. However only the kinetic
term of the M9-brane effective action has been constructed in the
literature \cite{BvdS,EL1}. Therefore,
in this section we derive the  KK-8A brane effective action by
performing a T-duality transformation in the NS-7B brane. The
T-duality takes place along a transverse direction, which in the
dual KK-8A brane plays the role of Killing direction. 
We interpret this direction as worldvolume because it is the one
inherited from the M9-brane in the double dimensional reduction 
explained before.

The result is the following effective action:
\begin{equation}
\label{IIAKK8}
\begin{array}{rcl}
&&S_{{\rm KK8A}} =
-T_{{\rm KK8A}}  \int d^8 \xi \,\,\, e^{-3\phi}
|k|^3 (1+e^{2\phi}|k|^{-2}(i_k C^{(1)})^2) \times \\
& & \\
& & \times
\sqrt{|{\rm det}(D_i X^\mu D_j X^\nu g_{\mu\nu}
-(2\pi\alpha^\prime)^2 |k|^{-2} {\cal K}^{(1)}_i
{\cal K}^{(1)}_j+\frac{(2\pi\alpha^\prime)|k|^{-1}e^\phi}
{\sqrt{1+e^{2\phi}|k|^{-2}(i_k C^{(1)})^2}}{\cal K}^{(2)}_{ij})|}\\
& & \\
& &
+ \,\, {1  \over 8!}(2\pi\alpha^\prime) T_{{\rm KK8A}} 
\int  d^8 \xi \,\, \epsilon^{i_1 \dots i_8}
\,\, {\cal K}^{(8)}_{i_1 \dots i_8} \, . \\
\end{array}
\end{equation}

\noindent The kinetic term is very similar to that of the Type IIA
KK-monopole (see \cite{BEL}), 
although now the factors in front of the square root
have different powers. 
The covariant derivative is the usual one for KK-monopoles:

\begin{equation}
DX^\mu=\partial X^\mu+Ak^\mu
\end{equation}

\noindent where $k^\mu \partial/\partial X^\mu=\partial/\partial z$ 
and $z$ is
the direction along which we T-dualized the NS-7B brane.
Therefore this direction becomes a Killing direction after T-duality.
The dependent gauge field is given, as usual, by:
$A=|k|^{-2}\partial X^\mu k_\mu$.
The field strengths ${\cal K}^{(1)}$ and ${\cal K}^{(2)}$ have the form:

\begin{equation}
\begin{array}{rcl}
{\cal K}^{(1)} &=& \partial\omega^{(0)}-\frac{1}{2\pi\alpha^\prime}
(i_k B^{(2)})   \, ,\\
& &\\
{\cal K}^{(2)} &=&2\partial\omega^{(1)}+\frac{1}{2\pi\alpha^\prime}
(i_k C^{(3)})-2{\cal K}^{(1)}(DX C^{(1)}) \, ,\\
\end{array}
\end{equation}

\noindent where $\omega^{(0)}$ and $\omega^{(1)}$ arise from the T-duality
transformations\footnote{Note that $\omega^{(1)}$ 
and $d^{(1)}$ (in (\ref{hachege})) have different gauge transformations.}

\begin{equation}
c^{(1)\prime}=-\omega^{(1)}\, ,\,\,\,
z^\prime=(2\pi\alpha^\prime)\omega^{(0)}\, ,
\end{equation}

\noindent and we also have

\begin{equation}
{\tilde {\cal F}}^\prime=-{\cal K}^{(2)}\, .
\end{equation}

\noindent ${\tilde \omega}^{(7)}$ is T-dual to ${\tilde c}^{(7)}$
in (\ref{Gtilde8}) and plays the role of tension of the KK-8A brane. 

\noindent Finally, the WZ term reads:

\begin{equation}
\label{IIAKK8WZ}
\begin{array}{rcl}
&&{\cal K}^{(8)} = 8\left\{ \partial 
{\tilde \omega}^{(7)} -\frac{1}{8(2\pi\alpha^\prime)}
(i_k N^{(9)})-(i_k N^{(8)})\partial\omega^{(0)}\right.
\\ & & \\ & &
+\frac72 (i_k N^{(7)})\left[ 2\partial\omega^{(1)}+
\frac{1}{2\pi\alpha^\prime}(i_k C^{(3)})+
\frac{2}{2\pi\alpha^\prime} (i_k B^{(2)})(DX C^{(1)})\right]
\\ & & \\ & &
-21(2\pi\alpha^\prime)(i_k B^{(6)}){\cal K}^{(1)}{\cal K}^{(2)}-
\frac{105}{4}(2\pi\alpha^\prime)(i_k C^{(5)})\left[{\cal K}^{(2)}
-4(DX C^{(1)}){\cal K}^{(1)}\right] {\cal K}^{(2)} 
\\ & & \\ & &
+70 DX^{\mu_1}DX^{\mu_2}DX^{\mu_3}C^{(3)}_{\mu_1\mu_2\mu_3}
\left[ (i_k C^{(3)})(i_k B^{(2)})-3(2\pi\alpha^\prime)^2
{\cal K}^{(1)}\partial\omega^{(1)}\right]{\cal K}^{(2)} 
\\ & & \\ & &
-35 DX^{\mu_1}DX^{\mu_2}DX^{\mu_3}C^{(3)}_{\mu_1\mu_2\mu_3}
(i_k C^{(3)})^2 \partial\omega^{(0)}
\\ & & \\ & &
-70 \left[ 2C^{(3)}(i_k B^{(2)})+3(i_k C^{(3)})B^{(2)}\right]
(i_k C^{(3)})(DX C^{(1)})\partial\omega^{(0)}
\\ & & \\ & &
+105 DX^{\mu_1}DX^{\mu_2}B^{(2)}_{\mu_1\mu_2}\left[(i_k C^{(3)})^2
{\cal K}^{(2)}-4(2\pi\alpha^\prime)^2 
(\partial\omega^{(1)})^3\right]
\\ & & \\ & &
\left. -\frac{7!}{3!}(2\pi\alpha^\prime)^3 A \partial\omega^{(0)}
(\partial\omega^{(1)})^3+\frac{7!}{3!}(2\pi\alpha^\prime)^3
\frac{e^{2\phi}k^{-2}(i_k C^{(1)})}{1+e^{2\phi}k^{-2}
(i_k C^{(1)})^2}({\cal K}^{(2)})^4 \right\} \, .\\
\end{array}
\end{equation}

\noindent This Wess-Zumino 8-form comes from 
(\ref{Gtilde8}) after T-duality along the transverse direction $z$.
The T-duality transformation rule for the field ${\tilde C}^{(8)}$
is:

\begin{equation}
\label{ctildeocho}
\begin{array}{rcl}
{\tilde C}^{(8)\prime}_{\mu_1\dots\mu_7 z}&=&
(i_k N^{(8)})_{\mu_1\dots\mu_7}+7(i_k N^{(7)})_{[\mu_1\dots\mu_6}
(C^{(1)}_{\mu_7]}-C^{(1)}_z \frac{g_{z\mu_7]}}{g_{zz}})\\
& & \\
& & +35(C^{(3)}_{[\mu_1\dots\mu_3}-3C^{(3)}_{[\mu_1\mu_2 z}
\frac{g_{z\mu_3}}{g_{zz}})C^{(3)}_{\mu_4\mu_5 z}
C^{(3)}_{\mu_6\mu_7 z} \\
& & \\
& & +70 [2C^{(3)}_{[\mu_1\dots\mu_3}B^{(2)}_{\mu_4 z}+ 
3C^{(3)}_{\mu_1\mu_2 z}B^{(2)}_{\mu_3\mu_4}]
C^{(3)}_{\mu_5\mu_6 z}(C^{(1)}_{\mu_7]}-C^{(1)}_z
\frac{g_{z\mu_7]}}{g_{zz}}) \, ,\\
\end{array}
\end{equation}

\noindent where $i_k N^{(8)}$ is the gravitational field to which 
the KK-6A brane couples minimally (see (\ref{KK6WZ})), and:

\begin{equation}
{\tilde C}^{(8)\prime}_{\mu_1\dots\mu_8}=
(i_k N^{(9)})_{\mu_1\dots\mu_8} \, .
\end{equation}

\noindent This last relation defines a new field $N^{(9)}$ (its
gauge transformation rule can be found in the Appendix),
which is electric-magnetic dual to the
mass parameter of the massive Type IIA supergravity
that is obtained by reducing the massive eleven dimensional
supergravity of \cite{BLO} along a direction different from the
Killing direction (see the Conclusions). As we have seen
this Killing
direction is also present in the worldvolume of the
KK-8A brane.
The field content of the KK-8A brane is that of the eight dimensional
vector multiplet, containing 1 vector field: $\omega^{(1)}$, and
2 scalars: $\omega^{(0)}$ and the transverse embedding scalar.

To finish this section let us analyze the soliton solutions 
in the worldvolume 
of the KK-8A brane.
These can be derived by T-duality from the configurations involving
the NS-7B brane, and also
from the configurations associated to the M9-brane
given in \cite{Mees,BGT}:

\begin{itemize}
\item The 1-form $\omega^{(1)}$ couples to a 0-brane soliton 
originated by the configuration $(0|{\rm D2},{\rm KK8A})$. This
is related by T-duality to $(0|{\rm D1},{\rm NS7B})$ and by
reduction to $(1|{\rm M2},{\rm M9})$ \cite{Mees,BGT}. This 1-brane
soliton is however realized as a 0-brane soliton in the effective
nine dimensional worldvolume of the M9-brane, given that one of
the worldvolume directions of the M2-brane is wrapped around the
isometric direction. Reduction along an ordinary worldvolume
direction of the M9-brane gives rise to $(0|{\rm D2},{\rm KK8A})$.
$\omega^{(1)}$ also couples magnetically to a 4-brane soliton,
associated to the configurations
$(4|{\rm NS5A},{\rm KK8A})$ and $(4|{\rm D4},{\rm KK8A})$.
The configuration $(4|{\rm NS5A},{\rm KK8A})$ is related by
T-duality to $(4|{\rm NS5B},{\rm NS7B})$ and is obtained from
the configuration \\$(5|{\rm M5}, {\rm M9})$, where the M5-brane
is wrapped around the Killing direction of the M9-brane. 
Therefore this soliton is realized as a 4-brane soliton in
the nine dimensional worldvolume of the M9-brane. On the other 
hand, the configuration $(4|{\rm D4},{\rm KK8A})$ 
is related by T-duality to $(4|{\rm D5},{\rm NS7B})$ and it is 
also obtained
from $(5|{\rm M5},{\rm M9})$, with the M5 embedded in the nine
dimensional worldvolume of the M9-brane and the reduction
taking place along a common worldvolume direction.

\item The 0-form $\omega^{(0)}$ couples magnetically to the 5-brane
soliton originated by the configuration $(5|{\rm NS5A},{\rm KK8A})$,
where the NS-5A brane is embedded in the KK-8A brane. This configuration
also arises from the reduction of $(5|{\rm M5},{\rm M9})$. 
It is related by T-duality to $(5|{\rm KK5B},{\rm NS7B})$.

\item The only physical embedding scalar can be associated to the boundary of a
KK-6A brane, where we make coincide the isometries of both,
the KK-6A and the KK-8A branes: $(5|{\rm KK6A},{\rm KK8A})$. 
This configuration arises as the dimensional reduction of
$(5|{\rm MKK},{\rm M9})$ along a direction of the
worldvolume of the M9-brane transversal to the MKK-monopole.

\item Since the KK-8A brane has the interpretation of an 8-brane in
10 dimensions, it is a domain wall, hence it is not expected to have a 
boundary. Thus we do not associate any physical meaning to the
7-form ${\tilde \omega}^{(7)}$.
\end{itemize}


\section{The IIB KK7-brane}
\label{IIB-KK-7-brane}


In this section we present the explicit effective action of an
exotic brane with two Killing isometries. This brane is
interpreted as a 7-brane in the Type IIB theory, and is non-perturbative
in the same sense as the branes that we have discussed previously,
since its tension also scales like $e^{-3\phi}$. It is related
to the KK-6A brane by T-duality along a transverse direction and
to the KK-8A brane by T-duality along a worldvolume direction.
This last connection allows to interpret this brane as the one
responsible for the mass of a massive 9 dimensional Type II supergravity 
containing a Killing direction. This theory should give rise to
an 8 dimensional massive supergravity with no Killing directions.

We construct the
worldvolume effective action by performing a
T-duality transformation in the KK-6A brane along a transverse
coordinate. The derivation starting from the KK-8A brane will 
also be discussed at the end of the section.
The result of these
duality transformations is a brane with 6 ordinary spatial directions
and two gauged isometries \cite{H1,PT},
which in the notation of \cite{H1} corresponds to the
$(6,1^2,1^3;3)$ exotic brane predicted by U-duality of M-theory
on an 8-torus. It has an effective tension:
\be
{\cal T}_{{\rm KK7B}} = {R_1^2 R_2^3 \over (2\pi)^6 g_s^3 l_s^{12}} \, ,
\ee
where $R_1, R_2$ are the radii of the two Killing directions.
We call this brane a KK-7B brane, since 
tracing back the origin of its two
Killing directions we find that one has its origin in the isometry of the MKK
monopole, which is interpreted as a transverse direction, whereas
the second Killing direction has its origin in the
isometry of the M9-brane, which is associated to a
worldvolume direction. Therefore, although
this brane contains only six ordinary spatial
directions, it is in fact interpreted as a 7-brane.

Since the exact worldvolume effective action is rather complicated
we have considered the case $B^{(2)}=C^{(2)}=0$.
Starting with the effective action of the KK-6A brane we perform
a T-duality transformation along a transverse direction
different from its Taub-NUT direction. This direction becomes a
Killing direction for the dual brane.
Denoting $h^\mu$ the Killing vector associated to the new isometry
the kinetic term in the dual effective action is given by:

\begin{equation}
\label{IIBKK7}
\begin{array}{rcl}
&&S_{{\rm KK7B}}=
-T_{{\rm KK7B}} \int d^7 \xi \,\,
e^{-3\phi} |h| |k^2 h^2-(k.h)^2| \times \\
& & \\
& &\hspace{-1.5cm}
\times\sqrt{|{\rm det}(D_iX^\mu D_jX^\nu g_{\mu \nu}
- (\alfa)^2 |h|^{-2} e^{\phi} {\cal K}_i^T M {\cal K}_j +
 (\alfa) |k^2 h^2 -(k.h)^2|^{-1/2} e^\phi {\cal G}^{(2)}_{ij})|}\, .
\end{array}
\end{equation}

\noindent Here the covariant derivatives contain the two Killing vectors
$k^\mu$ and $h^\mu$:

\begin{equation}
\label{covder1}
D X^\mu=\partial X^\mu +A^{(1)} k^\mu + A^{(2)} h^\mu\, ,
\end{equation}

\noindent and the two dependent gauge vector fields 
$A^{(1)}$, $A^{(2)}$ are defined 
as\footnote{In our notation: $(k.h)=-g_{\mu\nu}k^\mu h^\nu$.}:

\begin{equation}
\label{covder2}
A^{(1)}_i=\frac{h^2 k_i-(k.h) h_i}{k^2 h^2 -(k.h)^2}\, ,\,\,\,\, 
A^{(2)}_i=\frac{k^2 h_i-(k.h) k_i}{k^2 h^2 -(k.h)^2}\, .
\end{equation}

${\cal K}^T M {\cal K}$ is the SL(2,\R)-invariant 
combination\footnote{This combination appeared already in the 
effective action
of the Type IIB KK-monopole \cite{EJL}, which behaves as a singlet under 
S-duality transformations.}:

\begin{equation}
{\cal K}^T M {\cal K} = \left( \begin{array}{cc}{\cal K}^{(1)} 
& {\tilde {\cal K}}^{(1)} \end{array}
\right)  e^\phi \left( \begin{array}{cc}
e^{-2 \phi} + C^{(0)\, 2} &  C^{(0)} \\
C^{(0)} & 1 \end{array} \right)
\left( \begin{array}{c} {\cal K}^{(1)} \\ {\tilde {\cal K}}^{(1)}
\end{array} \right)
\, ,
\end{equation}
with
\begin{equation}
{\cal K}^{(1)} = \partial \omega^{(0)} \, ,\qquad
{\tilde {\cal K}}^{(1)} = \partial {\tilde \omega}^{(0)} \, .
\end{equation}

The two scalar fields $\omega^{(0)}$, ${\tilde \omega}^{(0)}$
transform as a doublet under S-duality:

\begin{equation}
\begin{array}{rcl}
\omega^{(0)} &\rightarrow &{\tilde \omega}^{(0)} \, ,\\
& &\\
{\tilde \omega}^{(0)} &\rightarrow & -\omega^{(0)}\, , \\
\end{array}
\end{equation}

\noindent and arise in the T-duality transformation as:

\begin{equation}
\label{scalars}
\begin{array}{rcl}
y^\prime &=& (2\pi\alpha^\prime)\omega^{(0)}   \, ,\\
& &\\
c^{(0)\prime} &=& -{\tilde \omega}^{(0)} \, .\\
\end{array}
\end{equation}

\noindent Note that here $y$ denotes the direction along 
which the T-duality
takes place, and should not be mistaken with the eleventh direction
in Section 2. Within M-theory,
recalling that $c^{(0)}$ (in the effective action of the KK-6A brane)
is related through
rescaling to the eleventh direction, we clearly see that
the S-duality transformation of the two scalars 
$\omega^{(0)}, {\tilde \omega}^{(0)}$ is induced from a modular
transformation in the 2-torus.

${\cal G}^{(2)}$ is given by:

\begin{equation}
{\cal G}^{(2)}=2\partial d^{(1)}+\frac{1}{2\pi\alpha^\prime}
(i_h i_k C^{(4)})+2(2\pi\alpha^\prime)\frac{(k.h)}{h^2}
\partial\omega^{(0)}\partial{\tilde \omega}^{(0)}\, ,
\end{equation}

\noindent and arises after T-duality as:

\begin{equation}
{\cal H}^{(2)\prime}={\cal G}^{(2)}\, .
\end{equation}

The Wess-Zumino term reads:

\begin{equation}
\label{IIBKK7WZ}
\begin{array}{rcl}
&&S^{{\rm WZ}}_{{\rm KK7B}} = \frac{1}{6!}(2\pi\alpha^\prime)
T_{{\rm KK7B}}\int d^7 \xi \,\, \epsilon \left\{ \partial 
{\tilde \omega}^{(6)} +\frac{1}{7(2\pi\alpha^\prime)}
(i_h i_k N^{(9)})+(i_h i_k N^{(8)})\partial\omega^{(0)}
\right.
\\ & & \\ & &
+(i_h i_k {\cal N}^{(8)})\partial {\tilde \omega}^{(0)}-
6(2\pi\alpha^\prime)(i_k i_h N^{(7)})\partial \omega^{(0)}\partial
{\tilde \omega}^{(0)}-6(i_k i_h {\cal N}^{(7)})\partial d^{(1)}+
\\ & & \\ & &
+5 DX^\mu DX^\nu DX^\rho (i_h C^{(4)})_{\mu\nu\rho}
\left[ \frac{1}{2\pi\alpha^\prime}(i_h i_k C^{(4)})^2+
6(2\pi\alpha^\prime)\partial d^{(1)}(2\partial d^{(1)}+ \right.
\\ & & \\ & &
\left. +\frac{1}{2\pi\alpha^\prime}(i_h i_k C^{(4)}))
\right]+120 (2\pi\alpha^\prime)^2 DX^\mu DX^\nu DX^\rho 
(i_k C^{(4)})_{\mu\nu\rho}\partial d^{(1)}\partial\omega^{(0)}
\partial {\tilde \omega}^{(0)}
\\ & & \\ & &
+30(2\pi\alpha^\prime)DX^\mu DX^\nu DX^\rho 
(i_k C^{(4)})_{\mu\nu\rho}(i_h i_k C^{(4)})\partial\omega^{(0)}
\partial {\tilde \omega}^{(0)}+
\\ & & \\ & &
\left. +360 (2\pi\alpha^\prime)^3 (\partial d^{(1)})^2 \partial
\omega^{(0)}\partial {\tilde \omega}^{(0)} A^{(2)}-
120 (2\pi\alpha^\prime)^2 A^{(1)} (\partial d^{(1)})^3
\right\} \, .\\
\end{array}
\end{equation}

\noindent The new fields $N^{(9)}$, $N^{(8)}$, ${\cal N}^{(8)}$,
arise from the 
T-duality of the fields $N^{(8)}$, $N^{(7)}$ in the Type IIA theory.
Namely, in our truncation:

\begin{equation}
\begin{array}{rcl}
(i_k N^{(8)\prime})_{\mu_1\dots\mu_7}&=&
(i_k i_h N^{(9)})_{\mu_1\dots\mu_7}+21
(i_k i_h {\cal N}^{(7)})_{[\mu_1\dots\mu_5}
(i_k i_h C^{(4)})_{\mu_6\mu_7]}\, ,\\
& & \\ 
(i_k N^{(8)\prime})_{\mu_1\dots\mu_6 y}&=&
(i_h i_k N^{(8)})_{\mu_1\dots \mu_6}\, ,\\
& & \\
(i_k N^{(7)\prime})_{\mu_1\dots\mu_6}&=&
(i_h i_k {\cal N}^{(8)})_{\mu_1\dots\mu_6}\, .\\
\end{array}
\end{equation}

\noindent Their gauge transformation rules 
can be found in the Appendix.

Similarly:

\begin{equation}
\begin{array}{rcl}
(i_k N^{(7)\prime})_{\mu_1\dots\mu_5 y}&=&
(i_k i_h N^{(7)})_{\mu_1\dots\mu_5}-
(i_k i_h {\cal N}^{(7)})_{\mu_1\dots\mu_5}
\frac{g_{yz}}{g_{yy}}\\
& & \\ & &
-5(i_k C^{(4)})_{[\mu_1\dots\mu_3}
(i_k i_h C^{(4)})_{\mu_4\mu_5]}-
5 (i_k i_h C^{(4)})^2_{[\mu_1\dots\mu_4}
\frac{g_{\mu_5]y}}{g_{yy}}\\
& & \\ & &
+\frac53 (i_h C^{(4)})_{[\mu_1\mu_2\mu_3}
(i_k i_h C^{(4)})_{\mu_4\mu_5]}\frac{g_{yz}}{g_{yy}}\, ,\\
\end{array}
\end{equation}

\noindent again in the same truncation. The field $N^{(7)}$ in 
the Type IIB theory is the electric-magnetic dual of the
Killing vector $k_\mu$ considered as a 1-form. This field is
the one to which the 
Type IIB Kaluza-Klein monopole couples minimally, when $k^\mu$
is taken along the Taub-NUT direction. Its gauge transformation
rule can be found in \cite{EJL}. In the action of
the KK-7B brane, with two Killing directions, there is another
7-form field, ${\cal N}^{(7)}$, electric-magnetic dual of the
second Killing vector, $h_\mu$. 
Analogously $N^{(8)}$ and ${\cal N}^{(8)}$ must be the 
electric-magnetic duals of the two scalars $k^2$, $h^2$. 
The field $N^{(9)}$ is dual to a mass parameter, which should
be the one of the nine dimensional Type II massive supergravity
with a Killing isometry
to which this 7-brane gives mass. 
${\tilde \omega}^{(6)}$ is T-dual to $\omega^{(6)}$ in (\ref{KK6WZ})
and plays the role of tension of the Type IIB KK7-brane. 

The effective action of the KK-7B brane is manifestly invariant
under the two local isometric transformations
generated by $k^\mu$ and $h^\mu$:

\begin{equation}
\delta X^\mu=-\sigma^{(1)}(\xi)k^\mu-\sigma^{(2)}(\xi)h^\mu\, .
\end{equation}

\noindent It is also easy to check that it is S-duality invariant.
The fields $N^{(9)}$, $N^{(7)}$ and ${\cal N}^{(7)}$ are S-self-dual
whereas $N^{(8)}$, ${\cal N}^{(8)}$ transform as a doublet. This can
be deduced from their origin in eleven dimensions, where S-duality
is realized as a modular transformation in $T^2$.

As we have mentioned before the KK-7B brane effective action can be
derived as well from the KK-8A brane, by performing a T-duality
transformation along a worldvolume direction. In this case the
T-duality mapping of the field strengths goes as follows:

\begin{equation}
\label{KK8AKK7B}
\begin{array}{rcl}
&&{\cal K}^{(2)\prime}_{ij}-2(2\pi\alpha^\prime)\frac{(k.h)}{k^2}
{\cal K}^{(1)\prime}_{[i}{\cal K}^{(2)\prime}_{j]\sigma}=
-{\cal G}^{(2)}_{ij} \\
& & \\ & &
{\cal K}^{(1)\prime}_i {\cal K}^{(1)\prime}_j+e^{2\phi}
{\cal K}^{(2)\prime}_{i\sigma}{\cal K}^{(2)\prime}_{j\sigma}=
e^{\phi}{\cal K}^T_i M {\cal K}_j \, .\\ 
\end{array}
\end{equation}

\noindent This connection with the KK-8A brane shows that 
one of the two Killing directions should be interpreted as a 
worldvolume direction, given that from this calculation
it is inherited from the worldvolume isometry of the KK-8A brane.

The worldvolume fields present in the effective action describe
soliton solutions in the worldvolume of the brane. In this case,
with two gauged isometries, the field strengths are of the form:

\begin{equation}
{\cal K}^{(p-1)}=(p-1)\partial\omega^{(p-2)}+
\frac{1}{2\pi\alpha^\prime}(i_h i_k C^{(p+1)})+\dots
\end{equation}

\noindent Therefore a $(p-2)$-form in the worldvolume couples to
a $(p-3)$-brane soliton realized as the boundary of a $p$-brane
wrapped around the two compact directions.
We find the following  soliton solutions in the 
worldvolume of the KK-7B brane:

\begin{itemize}
\item The 1-form $d^{(1)}$ 
couples to a 0-brane soliton originated by a D3-brane with two
of its worldvolume directions wrapped around the Killing directions
of the KK-7B brane: $(0|{\rm D3},{\rm KK7B})$. This configuration is
a singlet under S-duality and is related by T-duality to
$(0|{\rm D2},{\rm KK6A})$. 
A 0-brane soliton can also originate by a $(p,q)$ string wrapped
around one of the Killing directions of the KK7B. 
The $(0|{\rm D1},{\rm KK7B})$ configuration is related by T-duality
to $(0|{\rm D2},{\rm KK6A})$, whereas the $(0|{\rm F1},{\rm KK7B})$
is related to $(0|{\rm F1},{\rm KK6A})$. $d^{(1)}$ also couples
magnetically to a 3-brane soliton, realized as a $(p,q)$ 5-brane 
wrapped around the two Killing directions of the KK7B-brane.
These configurations are T-dual to $(3|{\rm NS5A},{\rm KK6A})$ and
$(3|{\rm D4},{\rm KK6A})$.

\item The field strengths of the two scalars
$\omega^{(0)}$ and ${\tilde \omega}^{(0)}$
involve the projection with the Killing vectors of the 2-forms
$C^{(2)}$ and $B^{(2)}$,  which we have not included. Accordingly, 
the doublet $(\omega^{(0)}, {\tilde \omega}^{(0)})$ will couple to the
4-dimensional boundary of a $(p,q)$ 5-brane which is wrapped
around one of the Killing directions. 
T-duality relates these configurations
with $(4|{\rm D6},{\rm KK6A})$ and $(4|{\rm KK5A},{\rm KK6A})$.

\item The embedding scalars are, as usual, associated to the
boundary of other Type II KK$p$ branes.
However, the branes involved in this case are some of the exotic
branes that we have not considered in this paper (see the
Conclusions), so we will omit the explicit configurations.

\item We find as well a domain-wall type of soliton, realized as
$(5|{\rm KK5B},{\rm KK7B})$, coupled to ${\tilde \omega}^{(6)}$,
and T-dual to $(5|{\rm NS5A},{\rm KK6A})$. 

\end{itemize}


\section{M-Theory Interpretation}
\label{MKK-domain-wall}


In this section we study the KK6A and NS7B branes from the point
of view of the M-theory Kaluza-Klein monopole solution, from which
they can be derived through reduction and 
duality\footnote{We thank Chris Hull, who has also considered 
7-brane solutions and their relations to M-theory, 
for a conversation on this point.}. We are able to 
show that if one imposses an extra isometry in this solution then it
can be written as a 2-torus bundle over a two dimensional transverse
space and that a modular transformation in the torus can be undone
by a change of coordinates in the transverse space. This symmetry
allows to connect the KK6A with the D6 brane and the NS7B with the
D7 brane. We discuss the consequences regarding the appearance of
these branes in the spacetime supersymmetry algebras.

Our starting point is the MKK-monopole solution, which
arises when at least one direction in spacetime is compact:
\be
ds^2_{MKK} = ds_{6,1}^2 - {1 \over H}( dz + {\vec V} \cdot d {\vec x})^2 
 - H (d {\vec x} \cdot d {\vec x}) \, .
\ee
It represents a 6-brane in a four dimensional transverse space, where
$z$ is a Taub-NUT isometry direction and
the function $H$ is harmonic in the three dimensional space
${\vec x} = (x_1,x_2,x_3)$ and is related to ${\vec V}$ by
${\vec \nabla}H = {\vec \nabla} \wedge {\vec V}$. This solution
is magnetically charged under the Kaluza-Klein vector
$i_k g$, where $k$ is the Killing vector associated to the isometry 
direction $z$ and $g$ denotes the spacetime metric.

If we impose an additional isometry in the transverse space
of the monopole\footnote{This
can be achieved by considering an infinite array of monopoles
\cite{BOL}.}
the harmonic function depends only on two of the three transversal 
coordinates, and we can gauge away 
two of the three components of the gauge vector ${\vec V}$.
We make the choice such that $H=H(x_1,x_2)$ and 
${\vec V} = (0,0,V_3)$, with $V_3=V_3 (x_1,x_2)$.
The monopole solution takes the form:
\be
ds^2_{MKK} = ds_{6,1}^2 - {1 \over H}| dz + (V_3 + i H) dx_3|^2 
 - H (dx_1^2 + dx_2^2) \, .
\ee
This metric can be interpreted as
a torus-bundle over a 2-dimensional base space:
\be
ds^2_{MKK} = ds_{6,1}^2 - {A \over \tau_2}
| dz + \tau (x_1,x_2)  dx_3|^2 
 - R^2(dx_1^2 + dx_2^2)
\label{MKK-F-theory}
\ee
where the fiber is a 2-torus with real periodic coordinates
$(z,x_3) \sim (z+1,x_3+1)$, a constant area modulus $A$, which can be 
introduced by rescaling $z$ and $x_3$, and a complex structure
$\tau = \tau_1 + i\tau_2 = V_3 + iH$. This defines $\tau$ as a
holomorphic function on the base space $x_1+ix_2$,
or anti-holomorphic in $x_2+ix_1$. 
Finally, $R^2 = H$ is the scale parameter of the base space.
Therefore
the MKK-monopole solution can be written as the product of two spaces
$\R^{6,1} \times B$, where $B$ is the torus-bundle.

The fibers have an $SL(2,\Z)$ symmetry:
\be
\tau \rightarrow  {a \tau + b  \over c \tau + d } \,\, ,\qquad 
\left( \begin{array}{c} z \\ x_3 \end{array} \right ) \rightarrow
\left( \begin{array}{cc} a & -b \\ -c & d \end{array} \right )
\left( \begin{array}{c} z \\ x_3 \end{array} \right ) \, .
\ee
Notice that $z$ and $x_3$ are now both isometry directions, 
$z$ is the Taub-NUT coordinate for a  
modulus $\tau$ and $x_3$ for a modulus $-{\tau}^{-1}$.

The solution (\ref{MKK-F-theory}), upon reduction over the torus fiber,
has two wrapping modes, the D6-brane and the KK-6A brane. The D6-brane can be 
obtained by reducing over the coordinate $z$
(re-absorbing $A$):
\be
\label{D6br}
\ba{rcl}
ds_{D6}^2 &=& \tau_2^{-{1 \over 2}} ds^2_{6,1} 
- R^2 \tau_2^{-{1 \over 2}} ( dx_1^2 + dx_2^2)
-   \tau_2^{{1 \over 2}} dx_3^2 \, ,\\
& &\\
e^\phi &=& \tau_2^{-{3 \over 4}} \, ,\qquad C^{(1)} = \tau_1 dx_3 \, .\\
\ea
\ee
On the other hand, 
the KK-6A brane is defined as the reduction over the coordinate $x_3$.
Taking $z$ along the M-theory circle, 
the KK-6A brane solution is given by \cite{PT}:
\be
\ba{rcl}
ds_{KK6A}^2 &=& ({\tau_2 \over |\tau|^2})^{-{1 \over 2}} ds^2_{6,1} 
- R^2 ({\tau_2 \over |\tau|^2})^{-{1 \over 2}} ( dx_1^2 + dx_2^2)
-   ({\tau_2 \over |\tau|^2})^{{1 \over 2}} dx_3^2 \, ,\\
& &\\
e^\phi &=& ({\tau_2 \over |\tau|^2})^{-{3 \over 4}} 
\, ,\qquad C^{(1)} = - {\tau_1 \over |\tau|^2} dx_3 \, ,\\
\ea
\ee

\noindent and is such that it is related to the D6-brane solution
(\ref{D6br}) by the transformation of the fiber-modulus:
\be
\tau \rightarrow - {1  \over \tau} \, .\qquad 
\ee

The coordinate transformation in the base space 
$(x_1,x_2) \rightarrow (y_1,y_2)$, where \cite{Groningen-Boys-2}:
\be
d(y_2 + i y_1) = \tau d(x_2 + i x_1) \, ,
\label{coordinate-transformation}
\ee
yields $\tau_2 /|\tau|^2$ as a harmonic function in $(y_1,y_2)$
such that the KK-6A brane solution takes exactly the same form as the
D6-brane. 
Therefore the modular transformation $\tau\rightarrow -1/\tau$
relating the two 6-brane solutions is equivalent to
a coordinate transformation in the base space. This transformation
connects two points of the base with fibers S-dual to each other
and therefore establishes the equivalence between the two 6-brane
solutions. This is the reason why
the D6-brane and the KK-6A brane do not have associated independent
central charges in the Type IIB spacetime supersymmetry algebra.

In the limit $A \rightarrow 0$, the MKK-monopole solution 
(\ref{MKK-F-theory}) yields the D7-brane solution with 
$\lambda = - {\bar \tau}$. Considering the compactification with a 
fiber modulus $-{\tau}^{-1}$ 
one obtains the NS-7B brane with $\lambda={\bar \tau}^{-1}$.
In particular the 7-brane solutions
(\ref{axion-dilaton-D7-brane}) and (\ref{axion-dilaton-NS7B-brane})
correspond to the following choice of $\tau$ as a section of the base space
$(x_1,x_2)$:
\be
\tau_{D7} = {1 \over 2\pi i} {\rm log} (x_2 - i x_1) \, ,\qquad
\tau_{NS7B} = - {1 \over \tau_{D7}} \, .
\label{sections-7-branes}
\ee
\noindent They can be written using the parameters of the
torus-bundle (\ref{MKK-F-theory}). For the D7-brane we have:
\be
\ba{rcl}
ds_{D7}^2 &=& \tau_2^{-{1 \over 2}} ( ds^2_{6,1} - dx_3^2 ) 
- R^2 \tau_2^{-{1 \over 2}} ( dx_1^2 + dx_2^2) \, ,\\
& &\\
e^\phi &=& \tau_2^{-{1}} \, ,\qquad C^{(0)} = - \tau_1 \, ,\\
\ea
\ee
and for the NS-7B brane:
\be
\label{solNS7}
\ba{rcl}
ds_{NS7B}^2 &=& ({\tau_2 \over |\tau|^2})^{-{1 \over 2}} 
( ds^2_{6,1} - dx_3^2)
- R^2 ({\tau_2 \over |\tau|^2})^{-{1 \over 2}} ( dx_1^2 + dx_2^2) \, ,\\
& &\\
e^\phi &=& ({\tau_2 \over |\tau|^2})^{-{1}} 
\, ,\qquad C^{(0)} = {\tau_1 \over |\tau|^2} \, .\\
\ea
\ee

\noindent The base space becomes the transverse space of the 7-branes
upon compactification and the S-duality relation between 
the two branes can be traced back to the properties of the
MKK-monopole interpreted as a torus-bundle. 
Namely, as we have seen above, 
the S-duality mapping is just 
a {\it gauge symmetry} in the  M-theory solution, 
in the sense that a coordinate transformation 
undoes the effect of the S-duality transformation. Thus,
a coordinate transformation in the space transversal to the 
7-branes relates both 7-brane solutions. As for the D6, KK6A brane 
solutions, this provides an
explanation for the fact that 
the D7 and the NS-7B branes do not appear as independent
branes in the Type IIB spacetime supersymmetry algebra.

The finite energy 7-brane solution of \cite{GGP} is obtained
from the monopole solution (\ref{MKK-F-theory})
by taking the modulus of the torus-bundle as:
\be
j(\tau) = {1 \over x_1 + i x_2} \, ,
\ee
such that we identify all solutions related by a modular transformation in 
the sections. This defines $\tau$ as a holomorphic function, and
yields the D7 (NS7B) brane solution in the weak (strong) coupling 
limit\footnote{After a $\pi/2$ rotation in the base space.}.

If we now impose one extra isometry in the harmonic function $H$, 
$H=H(x_1)$, it is not possible to
eliminate completely the dependence of the metric 
(\ref{MKK-F-theory}) on the coordinate $x_2$,
since according to the relation
between $H$ and ${\vec V}$, $V_3$ has to be a function of $x_2$ 
\footnote{See \cite{Groningen-Boys-2} 
for a discussion about this point.}. A particular type of this class
of configurations has been interpreted  in \cite{Hull-massive-strings}
as the compactification of M-theory giving rise to massive Type IIA
string theory\footnote{In this reference 
this solution was interpreted as an M9-brane. However,
we see here that it is just a Kaluza-Klein monopole 
with one extra isometry.}:
\be
\label{extra}
ds^2_{MKK} = ds_{6,1}^2  - R^2 (dx_1^2+dx_2^2)
- {A \over H}|dz+(mx_2+iH)dx_3|^2 \, .
\ee

\noindent This solution is obtained from (\ref{MKK-F-theory})
with the ansatz: $\tau = m x_2 + i H(x_1)$. We have substituted
$\tau_2=H$ to make more clear the connection with the solution
in \cite{Hull-massive-strings}. The condition 
${\vec \nabla}H = {\vec \nabla} \wedge {\vec V}$ fixes 
$H(x_1)=m|x_1|+{\rm const}$, so that (\ref{extra}) represents a
domain-wall at $x_1=0$ with two
different bundles at each side of the wall.

The limit $A \rightarrow 0$ yields the
Scherk-Schwarz reduction of the Type IIB theory with 
$\lambda = - {\bar \tau} (x_1,x_2)$ as ansatz for the axion-dilaton 
modulus. Taking the limit
$R \rightarrow 0$ as well, one obtains the T-dual theory, which is 
massive Type IIA string theory.
Thus in order to obtain a D8-brane solution from 
(\ref{MKK-F-theory}), one requires a bundle $B$ with monodromy
\be
\left( \begin{array}{cc} 1 & m \\ 0 & 1 \end{array} \right)
\label{mono}
\ee
in the zero volume limit $A \rightarrow 0$, $R \rightarrow 0$.
{}From the point of view of F-theory, the Scherk-Schwarz reduction
of Type IIB is a compactification on the
2-torus bundle over the circle in the $x_2$ direction (with
$H$ depending on the other transverse direction $x_1$) for fixed
area $A$ \cite{Hull-massive-strings}.

We know that T-duality along a transverse direction on the D7-brane 
gives the D8-brane \cite{BdRGPT}. 
This relation is also called {\it massive} T-duality and its
precise form is given by the kind of Scherk-Schwarz reduction
that is taken in the Type IIB side. This 
is determined by the monodromy (\ref{mono})
of the bundle $B$, so that it is possible to perform general Scherk-Schwarz
reductions with a monodromy matrix spanning the whole $SL(2,\Z)$
\cite{PT}.
These are related by T-duality to different massive modifications
of the Type IIA theory.

If we perform a T-duality transformation in the direction
$x_2$ transversal to the D7-brane, 
we obtain a D8-brane solution of the form
\be
\ba{rcl}
ds^2_{D8} &=& \tau_2^{- {1\over 2}} 
( ds^2_{7,1} - R^{-2} \tau_2 dx_2^2 )
- R^2 \tau_2^{- {1 \over 2}} dx_1^2 \, ,
\\& &\\
e^\phi &=& R^{-1} \tau_2^{- {3 \over 4}} 
\, ,\qquad C^{(1)} = (\tau_1 - m x_2)dx_2 \, .\\
\ea
\label{D8-solution}
\ee
For the T-duality the harmonic function $H$ must be taken 
as a function linear on
$x_1$, hence $\tau_1= V_3$ must be linear in $x_2$. We choose
$H=m|x_1|+{\rm const}$ and $V_3 = m x_2$, and therefore
$C^{(1)}=0$. 

Similarly, one can consider the T-duality of the NS-7B brane.
However, the metric associated to this solution depends,
through the modulus of $\tau$, on the $x_2$ coordinate 
in which we are performing the T-duality
transformation. Therefore we need to use the generalized Buscher
T-duality rules derived in \cite{PT}, which allow to treat this
case.
The result is the KK-8A brane \cite{PT}:

\be
\ba{rcl}
ds^2_{KK8A} &=& \tau_2^{{1\over 2}}
(ds^2_{7,1}-R^{-2}\tau_2^{-1}dx_2^2)-R^2\tau_2^{{1\over 2}}dx_1^2\, ,
\\& &\\
e^\phi &=& R^{-1}\tau_2^{{3 \over 4}}
\, ,\qquad C^{(1)}=0\, . \\
\ea
\ee

\noindent Both solutions, the D8 and the KK-8A branes, 
can be obtained from the M9-brane solution of \cite{BvdS}\footnote{We 
take $\epsilon=-1$.} by dimensional reduction. We can interpret this 
M9-brane as a torus-bundle over the one-dimensional transverse space
spanned by $x_1$:

\be
ds_{M9}^2 = R^{2 \over 3} ds_{7,1}^2
- R^{-{4\over 3}} \tau_2^{-1} |dz + \tau (x_1) dx_2|^2 - R^{8 \over 3} dx_1^2 
\, ,
\label{M9-torus-fiber}
\ee
with fiber-modulus $\tau (x_1) = iH(x_1)$ and $R^2=H$. 
In this case the area of the torus is also a function of $x_1$.
Reducing
along $z$ we recover the D8-brane solution whereas the reduction
along $x_2$ gives the KK-8A brane solution.
The relation between this two reductions is an S-transformation of 
the section $\tau$. Analogously to the MKK-monopole solution,
this modular transformation can be undone by a change of coordinates
in the transverse space. First, let us rescale the metric so that
$ds_{M9}^2$ takes the form:

\be
ds_{M9}^2=ds_{7,1}^2-{R^{-2} \over \tau_2} |dz+\tau dx_2|^2-
R^2 dx_1^2\, .
\ee

\noindent In this form it is clear that the area of the torus
$A=R^{-2}$, and therefore it is a function of $x_1$. The modular
transformation $\tau_2\rightarrow \tau_2^{-1}$ can be generated
by the change of coordinates:

\be
dx_1\rightarrow \tau_2 dx_1\, ,\qquad
dz\rightarrow \tau_2^{-1}dz\, ,\qquad
dx_2\rightarrow \tau_2^{-1}dx_2\, .
\ee

\noindent We see that in this case it is also necessary to perform 
a change of coordinates in the torus metric, due to the dependence
of its area on the transverse direction. Thus, a coordinate
transformation relates the two 8-brane solutions, from where we can
conclude that they represent the same physical object and therefore
should have associated a single central charge in the Type IIA
spacetime supersymmetry algebra.

Notice that in (\ref{D8-solution}) the most general solution
compatible with $\vec{\nabla}H=\vec{\nabla}\wedge\vec{V}$ corresponds 
in fact to $V_3=mx_2+ c$, with $c = {\rm constant}$,
which gives a D8-brane with 
$C^{(1)}=c\, dx_2$. This is indeed a solution to the massive
IIA supergravity, and is related by dimensional reduction 
to a non-diagonal M9-brane solution. This non-diagonal M9-brane
has the same form as in (\ref{M9-torus-fiber}) but with
fiber-modulus $\tau = c + iH(x_1)$. Moreover, it can be obtained from
(\ref{M9-torus-fiber}) by means of a T-transformation of the section,
which induces a constant real part for $\tau$.
This defines a 1-parameter family of (non-diagonal) M9-branes
interpreted as a torus-bundle over the one dimensional transverse
space. Furthermore, it is easy to check that the non-diagonal M9-brane can be
reduced to the usual diagonal one by redefining the $z$ coordinate.
Thus it is also a solution of the
massive 11-dimensional supergravity of \cite{BLO}.

The dimensional reduction of a non-diagonal M9-brane, 
with $\tau = c + iH$,
gives rise to a D8-brane with $C^{(1)}= c\, dx_2$,
and to a KK-8A brane:

\be
\ba{rcl}
ds^2_{KK8A} &=& ({H \over c^2+H^2})^{-{1\over 2}}
(ds^2_{7,1}-R^{-2}{H \over c^2+H^2}dx_2^2)-
R^2({H \over c^2+H^2})^{-{1\over 2}}dx_1^2\, ,
\\& &\\
e^\phi &=& R^{-1}({H \over c^2+H^2})^{-{3 \over 4}}
\, ,\qquad C^{(1)}=-{c \over c^2+H^2}dx_2\, . \\
\ea
\ee
 
\noindent This solution could be obtained as well by applying
the massive T-duality rules of \cite{PT} to the NS-7B brane
solution (\ref{solNS7}) with $\tau_1=mx_2+c$ and $\tau_2=H$.
It is related to the D8-brane solution with $C^{(1)}=c\, dx_2$
by an S-transformation of the section $\tau = c + iH$.

\section{Conclusions}


We have constructed the worldvolume effective action
associated to the NS-7B brane and checked that it transforms
as an $SL(2,\Z)$ triplet together with the D7-brane. The third field 
entering in this triplet transformation is the electric-magnetic
dual of the dilaton, which however does not have associated
an independent 7-brane solution: the D7 and NS-7B
brane solutions themselves are the ones carrying charge with
respect to this field.
We have clarified why the NS-7B brane
does not appear in the Type IIB spacetime supersymmetry
algebra. {}From one point of view this brane is charged with respect to
a field which is a combination of the duals of the axion and
dilaton (with coefficients depending as well on the axion and
dilaton), and therefore is not an independent field in the theory.
This function is however non-local and this is why this field
has to be introduced in the S-duality transformation rule.

{}From another point of view we have also shown that the D7 and
NS7B brane solutions and the D6 and KK-6A brane solutions
are related by a coordinate transformation
in the two dimensional transverse space which amounts to choosing
a different region to parametrize the $SL(2,\Z)$ moduli space.
This provides further evidence for the fact that both 7-branes
and 6-branes
represent the same physical object.
A similar kind of transformation relates the D8 and KK-8A brane
solutions, in this case in the one dimensional transverse space.

The relation

\begin{equation}
\label{r1}
d{\tilde C}^{(8)}=-C^{(0)}dB^{(8)}+|\lambda|^2dC^{(8)}\, ,
\end{equation}

\noindent between the field to which the NS-7B brane couples 
minimally and
the duals of the axion and dilaton translates under T-duality
into relationships between the fields to which the KK-6A and 
KK-8A branes couple minimally and the $C^{(7)}$, $C^{(9)}$
RR fields of the Type IIA theory. Namely:

\begin{equation}
\label{r2}
d(i_k N^{(8)})=(i_k C^{(1)})d(i_k B^{(8)})-[(i_k C^{(1)})^2+
e^{-2\phi}k^2]dC^{(7)}
\end{equation}

\noindent and:

\begin{equation}
\label{r3}
d(i_k N^{(9)})=(i_k C^{(1)})d(i_k B^{(9)})-[(i_k C^{(1)})^2+
e^{-2\phi}k^2]d(i_k C^{(9)})\, .
\end{equation}

\noindent The first expression gives a relation between the field 
$i_k N^{(8)}$, to which the KK-6A brane couples minimally, 
the RR 7-form and the T-dual of $B^{(8)}$, which we have
assumed to be again an 8-form: $B^{(8)\prime}_{\mu_1\dots\mu_7 z}
=B^{(8)}_{\mu_1\dots\mu_7 z}$. This 8-form
in the Type IIA theory can be given 
some interpretation in connection to the M8-brane of
\cite{EGKR,H1}. This BPS 8-brane solution of M-theory is required
by U-duality of M-theory on $T^8$, but is however not predicted
by the M-theory spacetime supersymmetry algebra. An
8-brane in M-theory requires the existence of a 9-form potential
to which it couples minimally. If we assume that the eleven 
dimensional background contains an isometry (say in the direction
$z$), along which we reduce
to go to the Type IIA theory, this 9-form potential could be the
electric-magnetic dual of the component of the
metric ${\hat g}_{zz}$. Then the 9-form would be 
interpreted as the dual of the
dilaton at strong coupling. When reducing to ten dimensions it would 
give rise to a 9-form and an 8-form, which would have associated an 8-brane
and a 7-brane respectively. The 8-form could be the $B^{(8)}$ field
above and the 9-form the $B^{(9)}$ field that appears in (\ref{r3}), 
and which comes as well from the T-duality of the field $B^{(8)}$ of
the Type IIB theory: 
$B^{(8)\prime}_{\mu_1\dots\mu_8}=B^{(9)}_{\mu_1\dots\mu_8 z}$.
These two fields can be interpreted in the Type IIA theory
as the dual of the dilaton, for $B^{(8)}$, and the dual of a mass
parameter, for $B^{(9)}$. A 7-brane and an 8-brane associated to
these fields are however not predicted
by the Type IIA spacetime supersymmetry algebra.
It is unclear to us at this point why this happens. 
One possibility is that
it is the D6, KK6A and D8, KK8A branes themselves the ones 
carrying charge with respect to the fields $i_k B^{(8)}$ and $B^{(9)}$
respectively, as happens with the 7-branes and the $B^{(8)}$ field in
the Type IIB theory.
The two relations (\ref{r2}) (\ref{r3}) suggest that the fields to which
the KK-6A and the KK-8A branes couple are not independent fields in the
theory, and this would explain why these branes do not appear in the
Type IIA spacetime supersymmetry algebra.  
One way to clarify further this issue would be to analyze the equations
of motion associated to the fields involved in these relations, i.e.
the RR 1-form, the dilaton and the $k^2$ field, to elucidate which
are the independent dual potentials in the theory. 
We hope to report
progress in this direction in the near future.
Finally, T-duality in (\ref{r2}) also establishes a relationship
between the field $(i_k i_h N^{(9)})$ to which the KK-7B brane
couples minimally and $i_h C^{(8)}$ (in the truncation 
$B^{(2)}=C^{(2)}=0$). As before, this could be the reason for the
non-occurrence of the KK-7B brane in the Type IIB spacetime
supersymmetry algebra. 

There are other exotic
branes required to fill up multiplets of BPS states of M-theory
on $T^d$ for $d\ge 8$ which we have not considered in this paper
\cite{EGKR,BOL,H1,OP}.
Some of them are very simply related to the branes that we have
studied, and we have indeed checked that the duality relations
that we mention below hold with the worldvolume effective
actions that we have constructed.
For instance T-duality of the KK-6A brane along a
worldvolume direction gives a 5-brane in Type IIB with two
Killing isometries, which in the notation of \cite{H1} corresponds
to the $(5,2^2;3)$-brane. This brane forms an S-duality doublet with
the $(5,2^2;2)$-brane obtained by T-duality of the Type IIA KK-monopole
along a transverse direction. Similarly,
T-duality of the KK-8A brane along the transverse direction
gives a 7-brane in Type IIB with two Killing directions, which
in the notation of \cite{H1} corresponds to a $(7,2^3;3)$ brane.
This brane forms an S-duality doublet with the $(7,2^3;4)$ brane
obtained from T-duality of the NS-9A brane along a worldvolume
direction \cite{H1}. Therefore in this case the two Killing
directions are interpreted as worldvolume directions, and 
the brane as a 9-brane. 

NS-9A branes seem to play a role in 
the description
of the Heterotic $E_8\times E_8$ as a non-perturbative orientifold 
of the Type IIA theory \cite{BEHHLvdS}. 
In this description the gauge group arises
in the form of Chan-Paton factors of (dimensionally reduced)
open M2-branes ending on the
8 (+8) M9-branes which are positioned on top of the 
two orientifold fixed-planes,
associated to the $I_{10}\Omega_M$
symmetry of M-theory (here $I_{10}$ is the
inversion of the eleventh direction and $\Omega_M$ reverses
the sign of the three form of eleven dimensional supergravity). 
In this description $\Omega_M$ is interpreted as the worldvolume
operation reversing one of the spatial directions of the M2-brane.
A T-duality transformation (along a worldvolume direction)
in this construction seems to indicate that the Heterotic SO(32) could be
obtained as an orientifold of F-theory on $T^2$ divided by the
symmetry $I_{10,11}\Omega_F$, with $\Omega_F=T\Omega_M T^{-1}$
($T$ denotes the T-duality transformation).
T-duality predicts that the $(7,2^3;4)$-branes above would
be the ones responsible for the gauge structure.
It would be interesting to explore this connection in more
detail and see in particular how the $\Omega_F$ symmetry could
be defined on the F-theory side.


\section*{Acknowledgements}


We would like to thank  A.~Arrizabalaga, E.~Bergshoeff, 
M.~de Roo, C.~Hull, N.~Obers and T.~Ort\'{\i}n
for useful discussions. E.E would like to thank the String Theory
group of Napoli University and especially F.~Pezzella, for hospitality.
He also thanks the people
of the Spinoza Institute for hospitality.


\appendix


\section{Appendix}


In this section we give the gauge transformation rules of the
new fields that couple to the effective actions of the branes
considered in the paper. The notation is that of \cite{EJL},
where the transformation rules of the fields coupled to 
D$p$-branes, NS5-branes and Kaluza-Klein monopoles can also
be found.

\begin{itemize}

\item The field to which the KK-6A brane couples minimally
transforms as:

\begin{equation}
\begin{array}{rcl}
\delta (i_k N^{(8)})&=&7\left\{ \partial (i_k \Omega^{(7)})-
15 \partial (i_k {\tilde \Lambda})(i_k C^{(3)})+
30 \partial \Lambda^{(2)} (i_k C^{(3)})^2 \right.
\\ & & \\ & &
\left. -20 C^{(3)} (i_k C^{(3)})\partial (i_k \Lambda^{(2)})
-(i_k N^{(7)})\partial\Lambda^{(0)} \right\} \, .\\
\end{array}
\end{equation}

\item The field $i_k N^{(9)}$ coupled to the KK-8A brane
transforms as:

\begin{equation}
\begin{array}{rcl}
&&\delta (i_k N^{(9)})=8\left\{ \partial (i_k \Omega^{(8)})
+21\partial (i_k \Sigma^{(6)})\left[ (i_k C^{(3)}+2(i_k B^{(2)})
(DX C^{(1)})\right] \right.
\\ & & \\ & &
+105\partial (i_k \Lambda^{(4)})(i_k C^{(3)})
\left[(i_k C^{(3)})+4(i_k B^{(2)})(DX C^{(1)})\right]
\\ & & \\ & &
+315 (i_k C^{(3)})^2\partial (i_k \Lambda^{(2)})
DXDXB^{(2)}
\\ & & \\ & &
+\frac{7!}{4}(i_k C^{(3)})\partial 
(i_k \Lambda^{(2)})B^{(2)}(i_k B^{(2)})(DX C^{(1)}) 
\\ & & \\ & &
-(2\pi\alpha^\prime)\left[ i_k N^{(8)}+7 (i_k N^{(7)})
(DX C^{(1)})+35DXDXDX C^{(3)} (i_k C^{(3)})^2 \right.
\\ & & \\ & &
\left. \left. +70 \left( 2C^{(3)}(i_k B^{(2)})+
3(i_k C^{(3)})B^{(2)}\right) (i_k C^{(3)})(DX C^{(1)})\right]
\partial (i_k \Lambda)\right\} \, .\\
\end{array}
\end{equation}

\item In the KK-7B brane $(i_h i_k N^{(9)})$ transforms as:

\begin{equation}
\begin{array}{rcl}
&&\delta (i_h i_k N^{(9)})= 7\left\{ \partial (i_h i_k \Omega^{(8)})
+6 (i_h i_k {\cal N}^{(7)})\partial (i_h i_k \Lambda^{(3)})\right.
\\ & & \\ & &
\left. -15 \partial (i_h \Lambda^{(3)})(i_h i_k C^{(4)})^2+
10 \partial (i_h i_k \Lambda^{(3)})(i_h C^{(4)})(i_h i_k C^{(4)})
\right\} \, .\\
\end{array}
\end{equation}

This is the correct gauge transformation rule in the 
truncation $B^{(2)}=C^{(2)}=0$. Also, in this truncation:

\begin{equation}
\begin{array}{rcl}
&&\delta (i_h i_k N^{(8)})=6\partial (i_h i_k \Omega^{(7)})
\\ & & \\ 
&&\delta (i_h i_k {\cal N}^{(8)})=6\partial 
(i_h i_k {\tilde \Omega}^{(7)})\, .\\
\end{array}
\end{equation}

$N^{(9)}$, $N^{(8)}$, ${\cal N}^{(8)}$ are now 9- and 8-forms 
in the Type IIB theory.

\end{itemize}


\end{document}